\documentclass[useAMS,usenatbib]{mn2e}
\usepackage[pdftex]{graphicx} 
\usepackage[hidelinks]{hyperref}
\usepackage{xcolor} 
\usepackage{titlesec}
\usepackage{lscape}
\usepackage{calc}
\usepackage{amsmath}
\usepackage{afterpage}

\makeatletter
\newcommand{\DESCRIPTION@original@item}{}
\let\DESCRIPTION@original@item\item
\newcommand*{\DESCRIPTION@envir}{DESCRIPTION}
\newlength{\DESCRIPTION@totalleftmargin}
\newlength{\DESCRIPTION@linewidth}
\newcommand{\DESCRIPTION@makelabel}[1]{\llap{#1}}%
\newcommand{\DESCRIPTION@item}[1][]{%
  \setlength{\@totalleftmargin}%
       {\DESCRIPTION@totalleftmargin+\widthof{\textbf{#1 }}-\leftmargin}%
  \setlength{\linewidth}
       {\DESCRIPTION@linewidth-\widthof{\textbf{#1 }}+\leftmargin}%
  \par\parshape \@ne \@totalleftmargin \linewidth
  \DESCRIPTION@original@item[{#1}]%
}
\newenvironment{DESCRIPTION}
  {\list{}{\setlength{\labelwidth}{0cm}%
           \let\makelabel\DESCRIPTION@makelabel}%
   \setlength{\DESCRIPTION@totalleftmargin}{\@totalleftmargin}%
   \setlength{\DESCRIPTION@linewidth}{\linewidth}%
   \renewcommand{\item}{\ifx\@currenvir\DESCRIPTION@envir
                           \expandafter\DESCRIPTION@item
                        \else
                           \expandafter\DESCRIPTION@original@item
                        \fi}}
  {\endlist}
\makeatother

\titleformat{\paragraph}
{\it\normalsize}{\theparagraph}{1em}{}
\titlespacing*{\paragraph}
{10pt}{3.25ex plus 1ex minus .2ex}{1.5ex plus .2ex}

\hypersetup{colorlinks,linkcolor={red!50!black}, citecolor={blue!50!black},urlcolor={blue!80!black} } 

\title[Galaxy groups in the low-redshift Universe]
{Galaxy groups in the low-redshift Universe}

\author[S.H. Lim et al.]{S.H. Lim$^{1}$\thanks{E-mail:
slim@astro.umass.edu}, 
H.J. Mo$^{1,2}$, 
Yi Lu$^{3}$, 
Huiyuan Wang$^{4}$,
Xiaohu Yang$^{5,6}$
\\ \\
$^{1}$Department of Astronomy, University of Massachusetts, Amherst MA 01003-9305, USA \\
$^{2}$Physics Department and Center for Astrophysics, Tsinghua University, Beijing 10084, China \\ 
$^{3}$Key Laboratory for Research in Galaxies and Cosmology, Shanghai Astronomical Observatory,
\\ ~~Nandan Road 80, Shanghai 200030, China \\ 
$^{4}$Key Laboratory for Research in Galaxies and Cosmology, Department of Astronomy,\\
~~University of Science and Technology of China, Hefei, Anhui 230026, China\\
$^{5}$Department of Astronomy, Shanghai Jiao Tong University, Shanghai 200240, China \\ 
$^{6}$IFSA Collaborative Innovation Center, Shanghai Jiao Tong University, Shanghai 200240, China \\ 
\ }

\begin{document} 

\date{Accepted ........ Received .......; in original form ......}

\pagerange{\pageref{firstpage}--\pageref{lastpage}}

\pubyear{2017}

\maketitle

\label{firstpage}

\begin{abstract} 
We apply a halo-based group finder to four large redshift surveys,
the 2MRS, 6dFGS, SDSS and 2dFGRS, to construct group catalogs 
in the low-redshift Universe. 
The group finder is based on that of 
Yang et al. but with an improved halo mass assignment so that it can be 
applied uniformly to various redshift surveys of galaxies. 
Halo masses are assigned to groups according to proxies based on 
the stellar mass/luminosity of member galaxies. The performances of 
the group finder in grouping galaxies according to common halos and 
in halo mass assignments are tested using realistic mock samples 
constructed from hydrodynamical simulations and empirical models
of galaxy occupation in dark matter halos. Our group finder 
finds $\sim 94\%$ of the correct true member galaxies for $90-95\%$ 
of the groups in the mock samples; the halo masses assigned 
by the group finder are un-biased with respect to the true halo masses, 
and have a typical uncertainty of $\sim0.2\,{\rm dex}$.
The properties of group catalogs constructed from the observational samples 
are described and compared with other similar catalogs in the literature.
\end{abstract} 


%%%%%%% SECTION 1
\section[intro]{INTRODUCTION}
\label{sec_intro}
%%%%% HJM revised

Grouping galaxies observed in a galaxy catalog into systems (clusters 
and groups) is a practice of long history. In the early attempts, clusters
of galaxies were identified based on optical photometric data, using the local
density contrast of galaxies in the sky as a proxy of spatial density and using  
distance estimates that are based on galaxy magnitudes. For example, 
\citet{abell58} constructed a catalog of about 2,700 clusters from the POSS 
plates using local galaxy surface number densities. A similar selection was 
used by \citet{abell89} to construct a catalog of 1,600 clusters
from the UKST plates. \citet{zwicky61} 
identified $9,133$ clusters in the northern celestial hemisphere using  
the POSS plates, and adopting a galaxy number density criterion that 
is relative to the immediate neighborhood. Because these catalogs 
are constructed from photographic plates and no redshift information is 
available for individual galaxies, they suffer severely in in-homogeneity, 
incompleteness, and projection effects.  
 
With the advent of large redshift surveys in 1980s, a lot of efforts 
were made to select galaxy clusters/groups on the basis of closeness 
of galaxies in redshift space. Although differing in details, many of these 
investigations have adopted the so-called friends-of-friends (FoF) method, 
which identifies galaxy systems as member galaxies that are linked by 
some adopted linkage criteria. For example, \citet{postman84} 
identified galaxy groups from the CfA redshift survey \citep{huchra83} 
by applying the FoF algorithm, developed by \citet{huchra82}, 
which uses two linking criteria, one on projected separation and
the other on redshift difference, to link galaxies. With modifications, 
the FoF algorithm  has been applied to various redshifts surveys of 
galaxies, including the Two Degree Field Galaxy Redshift 
Survey \citep[2dFGRS; e.g.][]{eke04}, the Two Micron All Sky Redshift Survey 
\citep[2MRS; e.g.][]{crook07}, and the Sloan Digital Sky Survey 
\citep[SDSS; e.g.][]{goto05, berlind06}. \citet{lavaux11} applied the FoF 
group finder to their own compilation combining the 2MRS, SDSS 
and Six Degree Field Galaxy Survey (6dFGS).

As high density regions in the galaxy distribution, clusters and groups of 
galaxies have been widely used to study the environmental dependence of the 
galaxy population and its evolution. For example, \citet{dressler80} found 
that the morphology of a galaxy is correlated with the local density of galaxies
in that the fraction of elliptical galaxies is higher
in regions of higher density. \citet{butcher78, butcher84} studied
the galaxy populations in rich and compact clusters at redshifts of $\sim0.4$ 
and found that the ratio of blue galaxies is higher than that in nearby clusters of 
similar richness and morphology, implying a strong recent evolution 
in galaxy color. Galaxy systems have also been assumed to be associated with 
dark matter halos.  
In the 1930s, Zwicky studied the motion of galaxies within the 
Coma Cluster and found that the total mass of the cluster estimated using the 
virial theorem is more than $100$ times higher than that estimated from 
the total luminosity of member galaxies. This is the first evidence 
for the presence of a large amount of non-luminous (dark) matter in 
clusters of galaxies. While the result was not widely accepted at the 
time, subsequent observations based on galaxy velocity dispersion, 
X-ray emission and gravitational lensing effects have provided 
indisputable evidence that galaxy clusters and groups are all associated 
with massive dark matter halos. Indeed, even isolated galaxies are also found  
to be embedded in massive halos, as inferred from their rotation curves 
and velocity dispersion of stars.  

Theoretically, the current $\rm{\Lambda CDM}$ model predicts that all galaxies 
form and evolve in dark matter halos. These halos are virialized 
clumps of dark matter that form in the cosmic density field through gravitational
instability \citep[see][for a review]{mo10}.
Therefore, galaxy systems, if selected properly so as to represent halos,
can be used to study how galaxies form and evolve in dark matter halos.  
Furthermore, since dark matter halos are simple but biased tracers of the 
underlying mass density field \citep[e.g.][]{mo96}, galaxy systems so 
selected can also be used to study the structure and evolution of the mass 
density field in the universe. In particular, as shown in \citet{wang09},
a well-defined group sample can be used to reconstruct the cosmic density 
field, which, in turn, can be used to reconstruct the initial conditions 
from which the observed structures form and evolve \citep[][]{wang13, wang14, 
wang16}. 

A key in using galaxy systems as a proxy of the dark halo population
is a group finder that can group galaxies according to common dark matter halos. 
The widely adopted FoF algorithm is not optimal for the purpose. 
More recently, \citet{yang05} (Y05 hereafter) developed a 
halo-based group finder, which identify groups based on dark matter halo 
properties, such as mass and velocity dispersion, expected from the 
CDM cosmogony. This halo-based group finder has been extensively tested 
using mock galaxies from simulations and found to perform much better than 
the traditional FoF algorithm, particularly in identifying poor systems. 
The group finder of Y05 has been applied to redshift surveys such as the 2dFGRS 
\citep[e.g.][]{yang05}, the SDSS \citep[e.g.][]{weinmann06, yang07}, 
and the 2MRS \citep[e.g.][]{lu16}. Similarly, \citet{duarte15} adopted 
an iterative group membership assignment algorithm but in a probabilistic way 
using galaxy distribution statistics extracted from N-body simulation. 
An important step in the halo-based 
group finder is the use of a halo-mass proxy to assign halo masses to tentative 
groups/clusters in the grouping process. Y05 suggested the use of the ranking 
of the total luminosity of galaxies that have luminosities above a certain value 
as the proxy of halo mass, and this mass proxy was adopted in the SDSS and 
2dFGRS group catalogs mentioned above. However, this mass proxy may not 
suitable for shallower surveys, such as the 2MRS, where many systems 
contain only a small number of galaxies. In order to overcome this 
limitation, \citet{lu16} (L16 hereafter) proposed a ``GAP correction" 
method, in which the luminosity/stellar mass of the most luminous/massive 
member is combined with the ``GAP" to form a mass proxy, 
where ``GAP" is defined to be the difference in luminosity/stellar mass between 
the most luminous/massive member and $n$-th most luminous/massive member. 
Using a 2MRS mock sample, L16 found that the GAP method yields a typical 
dispersion of $\sim0.3\,{\rm dex}$ in the estimated masses for galaxy systems 
of a given true halo mass. 

In this paper, we modify the group finders of Y05 and L16, paying
particular attention to the extension of the methods to poor
systems, such as groups containing one member or a small number of members, 
in a uniform way. 
We use mock samples constructed from numerical simulations and 
an empirical model to calibrate the halo mass proxies and to test the performances
of the group finder under different sample selections. As we will see below, 
our modified group finder not only gives more accurate halo mass estimates
for groups than the original group finders, it also enables us to uniformly 
extend the group samples to systems with halo masses that are about an order of magnitude 
lower than in the existing group catalogs. We apply our group finder  
to a number of redshift surveys in the local universe, including the 2MRS, 
the 6dFGS, the updated release of SDSS and the 2dFGRS. As mentioned above, 
group catalogs have been constructed from some of these catalogs with various 
group finders. Our goal here is to extend, update, and add values to, these 
catalogs by providing group samples that are uniformly selected from improved 
data using improved methods.  

The outline of this paper is as followings. In Section~\ref{sec_data}, we 
describe the observational data to which we apply our group finder, and 
the simulation that we use to calibrate and test the group finder. 
Section~\ref{sec_gfinder} explains in detail the group finder and how to test and
calibrate a variety of halo-mass proxies using mock galaxies. In Section~\ref{sec_test}, 
we apply the group 
finder to the mocks of the same sample selection as the observational data,  
and assess its performance by comparing halo masses, membership assignments, 
and global completeness between the constructed mock groups and the simulations. 
In Section~\ref{sec_gcatalog}, we apply the group finder to real 
observations, construct our group catalogs, describe their basic properties 
and how to use them, and make comparisons with other catalogs in the
literature. Finally, we summarize our results and discuss
applications of the catalogs in Section~\ref{sec_summary}.

%%%%%%% SECTION 2
\section[data]{OBSERVATIONAL DATA AND MOCK SAMPLES}
\label{sec_data}

\begin{figure}
\includegraphics[width=1.0\linewidth]{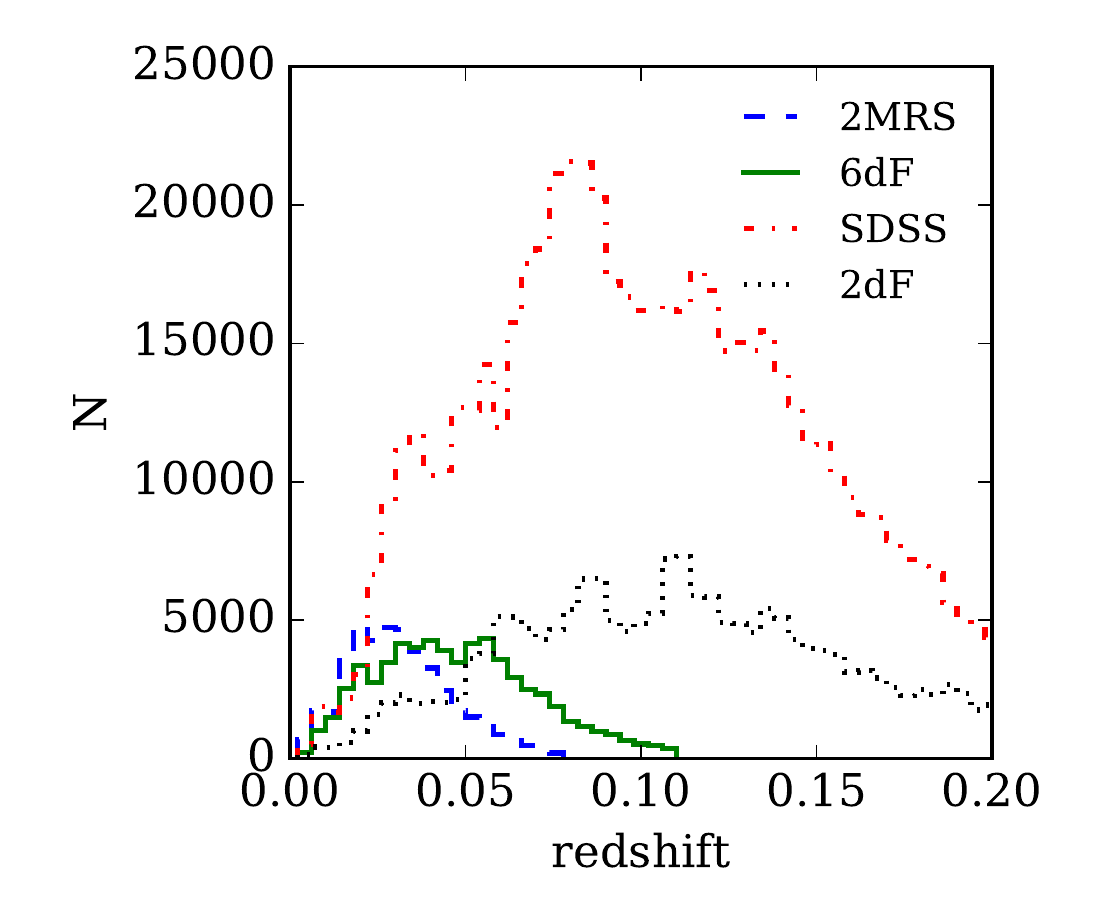}
\caption{The redshift distributions of galaxies in the four samples 
we use to identify galaxy groups. The bin size of the histograms 
is $\Delta z=0.004$.}
\label{fig_z_dist}
\end{figure}

\begin{figure*}
\includegraphics[width=0.75\linewidth]{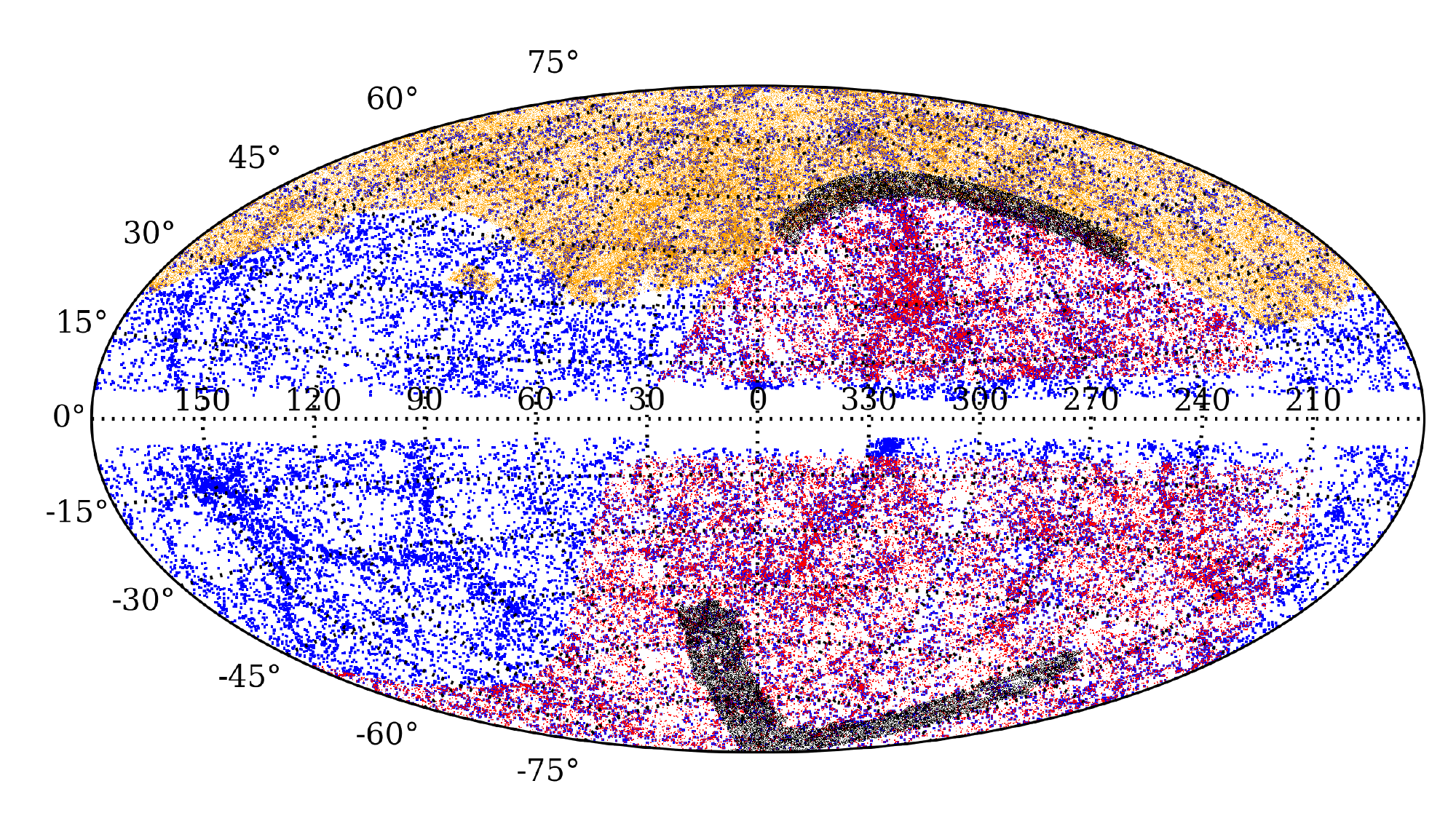}
\caption{The galaxy distributions in Galactic 
coordinates (Aitoff projection) of the 2MRS (blue), 6dFGS (red), 
SDSS (orange), and 2dFGRS (black) samples. }
\label{fig_proj_dist}
\end{figure*}

In this section, we describe in detail the galaxy samples we use 
to construct our group catalogs. Since our goal is to provide well-defined 
group catalogs in the local universe, we decide to use all major redshift 
surveys at low redshift ($z<\sim0.2$) that are publicly available.
A brief summary of our sample selections is given in Table~\ref{tab_sample}. 
The redshift distributions of these samples 
are shown in Figure~\ref{fig_z_dist}, and their sky coverages 
are plotted in Figure~\ref{fig_proj_dist}. 

%%%%%%%%%%% figure1

%%%%%%%%%%% figure2

%%%% Table 1 %%%%
\begin{table*}
 \renewcommand{\arraystretch}{1.5} 
 \centering
 \begin{minipage}{105mm}
  \caption{A summary of galaxy samples.}
  \begin{tabular}{ccccc}
\hline
Sample & Sky Coverage & 
Depth\textsuperscript{\footnote{Upper limit of redshift in CMB rest-frame.}}
& Magnitude limit &  No. of Galaxies\textsuperscript{\footnote
{The numbers in parentheses are for the extended (the catalogs with `+') 
catalogs.}} \\
 & ($\%$) & ($z$) & (mag) & \\
\hline
\hline
2MRS & $91\%$ & $0.08$ & $K_s\leq 11.75$ & $43,249\ (44,310)$ \\

6dFGS & $40\%$ & $0.11$ & $K_{s,tot}\leq 12.5$ & $62,987\ (73,386)$ \\

SDSS & $21\%$ & $0.2$ & $r\leq 17.77$ & $586,025\ (600,458)$ \\

2dFGRS & $3.3\%$ & $0.2$ & $b_J\leq 19.45$ & $180,967\ (189,101)$ \\

\hline
\vspace{-2mm}
\end{tabular}
\textbf{Notes.}
\vspace{-5mm}
\label{tab_sample}
\end{minipage}
\end{table*}
%%%%%%%%%%%%%%%%%%%%%%%%%%%%%%%%%%%%%%%%%%%%%%%%%%%%%%%%%%%%%%%%%%%%%%%%%%%%%%

\subsection{The 2MRS catalog} 
\label{ssec_2MRScat}

 Our first galaxy sample is selected from the 2MASS Redshift Survey 
\citep[2MRS;][]{huchra12}, which is based on the Two Micron 
All Sky Survey \citep[2MASS;][]{skrutskie06}. 2MASS covers 
$\sim91\%$ of the entire sky in the near-infrared $J$, $H$, 
and $K_s$ bands. Because of the reduced dust extinction 
in the NIR, 2MASS is an almost uniform survey down to the 
magnitude limit $K_s\leq 13.5$, except in the region within
$\pm 5^\circ$ of the Galactic plane (the ``zone of avoidance'' or ZoA).
The extended source catalog (the 2MASS XSC) contains $\sim10^6$ objects.
The 2MRS attempted to obtain redshifts, either from its own 
observation or from other data bases, for $45,086$ sources of 
the 2MASS XSC that meet the following criteria:
\begin{enumerate}
    \item $K_s\leq 11.75$ mag and detected at $H$,
    \item $E(B-V)\leq 1$,
    \item $|b|\geq 5^\circ$ for $30^\circ\leq l \leq 330^\circ$; 
             $|b|\geq8^\circ$ otherwise,
\end{enumerate}
where $b$ is the Galactic latitude, and $E(B-V)$ is the extinction
based on the dust map of \citet{schlegel98}.  
As shown in \citet{huchra12}, to the magnitude limit $K_s= 11.75$
the completeness does not change significantly within the region 
specified by criterion (iii). Of the $45,086$ sources,
2MRS rejected a small fraction that is of galactic origin, 
only partially detected, or not clearly detected due to 
contamination. This leaves a total of $44,599$ galaxies. 
The details about the selection can be found in \citet{huchra12}
 and its appendix. For the $44,599$ galaxies, 2MRS 
eventually obtained redshifts for $43,533$ systems, achieving 
a completeness of about $97.6\%$. These include $11,000$ galaxies
measured by the 2MRS team, $7,069$ galaxies with redshifts 
from SDSS, $11,763$ from 6dFGS DR3, $12,952$ from the 
NASA Extragalactic Database (NED), and $749$ from J. Huchra's 
personal compilation (ZCAT). For objects with redshifts from more 
than one source, the preference was given in the order of 2MRS, 
SDSS, 6dF, NED, and ZCAT. 

For the $1,066$ galaxies that do not have redshifts from the 2MRS, 
we either adopt redshifts of their nearest neighbors or 
use those given by the 2MASS Photometric Redshift catalog 
\citep[2MPZ;][]{bilicki14}. The 2MPZ uses the optical, NIR, and mid-IR
photometry from SuperCOSMOS, 2MASS, and WISE respectively, to obtain 
photometric redshifts for about 1 million galaxies, by employing an 
artificial neural network approach trained with several redshift surveys. 
The photometric redshifts obtained have a typical error of $12\%$.
We first assign redshifts of the nearest neighbors ($z_{NN}$) to all galaxies 
without the spectroscopic redshift. Then, if a galaxy also has a redshift from 
the 2MPZ ($z_{pho}$) and $z_{pho}$ differs from $z_{NN}$ by more than $12\%$, 
we assign $z_{pho}$ as the redshift of the galaxy instead of $z_{NN}$. 

With all these, we obtain redshifts for $288$ additional galaxies from their 
nearest neighbors, and redshifts for $778$ galaxies from the 
2MPZ, thus assigning redshifts to all the $44,599$ galaxies. Because of the 
uncertainties in the nearest-neighbor and photometric redshifts, we will
provide two separate catalogs: the first is constructed 
from the sample of galaxies that all have 2MRS redshifts;
the second uses all galaxies that have 2MRS redshifts, $z_{pho}$, or $z_{NN}$.
The latter will have a flag that shows source of redshifts for each galaxy, 
as well as the separation to the nearest neighbor for galaxies with $z_{NN}$ so 
that a user can decide an uncertainty that may be allowed to suit his/her
scientific goal. For convenience, we refer to the 1st catalog as 2MRS and the 2nd as 
2MRS+. 

Our sample also contains a number of refinements  appropriate for our 
purpose. First, we correct all the redshifts (radial velocities) 
of galaxies to the CMB rest-frame. To do this, we assume that 
the heliocenter is moving with a velocity of $368{\rm\ km/s}$ 
towards $(l,b)=(263.85^\circ, 48.25^\circ)$ with respect to the CMB
\citep[][]{bennett03}. Second, we only use galaxies with corrected redshifts
$z\leq0.08$, which eliminates about $1\%$ of the galaxies from the sample. 
The final numbers of galaxies are then $43,249$ and $44,310$ for 2MRS and 
2MRS+, respectively. As an example, the redshift and sky distributions of 
the galaxies in 2MRS are shown in Figures~\ref{fig_z_dist} and \ref{fig_proj_dist}, 
respectively, and a brief summary of the samples is given in 
Table~\ref{tab_sample}.

We use the extinction-corrected $K_s$ isophotal magnitudes from the 2MRS. 
The extinction-correction accounts for dust extinctions of the Milky Way
relying on the dust map by \citet{schlegel98}. 
We use WMAP9 cosmology to convert apparent magnitudes to absolute 
magnitudes in a bandpass $Q$ as followings:
%%%%%%%%%%%%%%%%%%%%%%%%%%%%%%%%%%
\begin{eqnarray}\label{eq_}
M_Q = m_Q + \Delta m_Q - DM(z) - {\cal K}_Q(z) - {\cal E}_Q(z) - {\cal S}_Q(z)
\end{eqnarray}
%%%%%%%%%%%%%%%%%%%%%%%%%%%%%%%%%%%
where $\Delta m_Q$ is the zero-point correction from the survey 
photometric system to the Vega system (or the AB system for surveys
introduced later in this section that use the AB system),
which is $0.017$ for the 2MASS $K_s$-band filter \citep[][]{cohen03},
$DM(z)$ is the distance modulus at $z$, ${\cal K}_Q(z)$ and ${\cal E}_Q(z)$ 
are the $K$- and evolution- corrections at redshift $z$, respectively,
and ${\cal S}_Q(z)$
corrects for the effect of decreasing aperture size within which flux 
is integrated with increasing redshift due to dimming of surface brightness. 
The term ${\cal S}_Q(z)$ is not needed when using extrapolated total 
magnitudes. We follow \citet{lavaux11} to model ${\cal K}_Q(z)$, 
${\cal E}_Q(z)$, and ${\cal S}_Q(z)$, and correct the values of $M_Q$ of 
individual galaxies to redshift $z=0.1$. For nearby 
galaxies that have negative recession velocities 
in the CMB rest-frame (a total of $25$ galaxies in the 2MRS catalog), 
we adopt distances from `EDD distances' available at the Extragalactic
Distance Database \citep[EDD;][]{tully09} to calculate the absolute magnitudes. 
These distances, however, are not used in identifying galaxy systems via the 
group finder, as our group finder works in redshift space.
In the cases where we do not find matches from the EDD (a total of $5$ galaxies
in the 2MRS catalog), 
we assign the distances of their nearest neighbors that have EDD 
distances available. Later in \S\ref{sec_gcatalog} where we 
construct group catalogs, we estimate stellar mass using the mean relation 
between stellar mass and $K_s$-band luminosity from the simulation described
in \S\ref{ssec_mock}.

\subsection{The 6dFGS catalog} 
\label{ssec_6dFGScat}
%%%%% HJM revised
%%%%% slim revised

Our second sample is selected from the 6dF Galaxy Survey 
\citep[6dFGS;][]{jones04, jones05}. Specifically we use the 
6dFGS Data Release 3 \citep[6dFGS DR3;][]{jones09}, the final redshift 
release of the survey. As 2MRS, the 6dFGS is based mainly on the 
$K_s$-selected 2MASS, but is deeper, with a magnitude limit of 
$K_{s,tot}= 12.65$ mag, where $K_{s,tot}$ is the total 
magnitude from the 2MASS. Note that the magnitudes are 
corrected for foreground dust extinction, as mentioned above.  
As shown in \citet{mcintosh06}, these 
magnitudes are robust against uncertainties in surface brightness. 
The survey has a sky coverage of $\sim41\%$ in the southern hemisphere. 

According to \citet{jones09}, the final 
6dFGS catalog contains $126,754$ unique redshifts from 
their own observations, 
$563$ redshifts from the SDSS, $5,210$ redshifts from the 2dFGRS, 
and $9,042$ redshifts from the ZCAT. For their own observations, the catalog
contains only the spectra with quality parameters $Q=3$ and $Q=4$, 
which are appropriate for scientific analysis according to \citet{jones09}.
Redshifts with $Q=3$ and $Q=4$ have typical uncertainties of $55\, {\rm km/s}$
and $45\,{\rm km/s}$, respectively \citep[see][for details]{jones09}.
However, the 6dFGS has poorer coverage in some regions, 
such as those toward the Large Magellanic Cloud (LMC) and the 
South Pole. This can affect the performance of our group finder. 
To reduce this effect we select a shallower sample,  
using the 2MASS Extended Source Catalog (2MASS
XSC) with a flux limit of $K_{s, tot}=12.5$ mag as an input catalog. 
The 2MASS XSC with this flux limit contains $75,098$ entries. 
For galaxies that have spectroscopic redshifts from the 6dFGS catalog, 
we assign the 6dF redshift. We also find redshifts for $1,533$ galaxies from 
the 2M++ galaxy redshift catalogue \citep[2M++;][]{lavaux11}, which are 
originally from the NED. For galaxies without spectroscopic redshifts available, 
we assign redshifts of their nearest neighbors or from the 2MPZ 
in the same way as for the 2MRS described above.  
Of all the 2MASS XSC galaxies, $62,929$ have
redshifts from the 6dFGS DR3, $1,533$ redshifts from the 2M++, 
$3,354$ redshifts from the nearest neighbor, and $7,282$ redshifts from 
the 2MPZ. This, of course, corresponds to $100\%$ redshift completeness. 
In the end, we will provide two separate catalogs: one constructed 
using only galaxies with spectroscopic redshifts, and 
the other using all galaxies, including the ones with nearest-neighbor 
and 2MPZ redshifts. For convenience, we refer to the 1st catalog as
6dFGS, and the 2nd catalog as 6dFGS+. 

For our analysis, we correct all the radial velocities (redshifts) 
to the CMB rest-frame, as we did for the 2MRS, and we only use 
galaxies with corrected redshift $z\leq 0.11$. 
This leaves $62,987$ and $73,386$ galaxies in our final 6dFGS
and 6dFGS+ samples, respectively. 
The redshift distribution and a summary of the final samples are given 
in Figure~\ref{fig_z_dist} and Table~\ref{tab_sample}, respectively.  

The absolute magnitudes of individual galaxies are again calculated
using equation (1). The same $K$-, $E$-, and surface brightness
corrections as those for the 2MRS are used to correct the $K_s$-band 
magnitudes to $z=0.1$. For galaxies with negative recession velocity, 
we again use the EDD distances to compute their luminosities. 
We approximate stellar mass using the mean relation 
between stellar mass and $K_s$-band luminosity obtained from the 
simulation described in \S\ref{ssec_mock}.

\subsection{The SDSS catalog}
\label{ssec_SDSScat}
%%% HJMO have revised the text
%%% slim revised

Our third sample is selected from the Sloan Digital Sky Survey Data 
Release 13 \citep[SDSS DR13;][]{albareti16}. DR13 is the first data release 
of the fourth phase of the Sloan Digital Sky Survey (SDSS-IV) and is 
built upon prior releases. It includes updated data for 
the SDSS Legacy Survey, which is a magnitude limited redshift 
survey completed in SDSS-II, as well as objects from the Baryon 
Oscillation Spectroscopic Survey \citep[BOSS;][]{dawson13}, the selection 
of which barely overlaps with that of the legacy survey. 
The main part of the SDSS Legacy Survey was already released in DR7, and 
remained more or less steady through DR12. 
Significant changes were made to photometric calibration in the DR13, 
including updated zero points and flat-fields in the $g$, $r$, $i$, and 
$z$ bands from the hypercalibration procedure of \citet{finkbeiner16}. 
These affect all photometric quantities of the galaxies in the Legacy Survey. 
In addition to the updated photometry calibration,
another significant improvement in DR13 relative to, for example, DR7 is that 
some of the fiber-collision galaxies in DR7 have their redshifts measured 
in DR13. The SDSS spectrograph used for the Legacy Survey did not allow 
two fibers to be positioned within $55\arcsec$, and so no spectroscopic 
measurement was available for galaxies that have close neighbors  
within the fiber separation, the so-called `fiber-collision' galaxies. 
Many of the `fiber-collision' galaxies \citep[$\sim 60\%$, e.g.][]{guo15}
have been measured spectroscopically in the later data releases through 
DR13. The Legacy Survey covers approximately $\sim 23\%$ of the sky, and is 
complete to an extinction-corrected Petrosian magnitude of $17.77$ mag 
in the $r$-band. 

From the full photometric catalog of DR13, we select all objects that are in the 
Legacy Survey region and identified as galaxies (${\rm type} =3$) brighter
than the $r$-band magnitude limit of $17.77$. We take the photometric quantities 
only from the primary observation in the cases where an object was observed multiple 
times (${\rm mode}=1$). We also get rid of galaxies in the Southern Galactic Cap, 
as its narrow angular boundary makes our group finder unreliable for
many systems close to the boundary. Note that these selections may include 
some of the BOSS galaxies that pass the selection criteria. The selections leave a 
total of $638,191$ entries, of which $16,251$ galaxies do not have redshifts 
for reasons such as fiber-collisions, broken or unplugged fibers, bad spectra, 
or poor fit to models. Of the $621,940$ galaxies that have redshifts, $20,780$ 
($\sim 3.3\%$) are BOSS galaxies in the Legacy region.  

For the $16,251$ galaxies without SDSS redshifts, 
we find redshifts from other sources: the 2dFGRS, 6dFGS, 
the Korea Institute for Advanced Study Value-Added Galaxy Catalog 
\citep[KIAS VAGC;][]{choi10}, a complementary galaxy sample in the LAMOST 
Survey \citep[][]{luo15, shen16}, the nearest neighbors, or the 2MPZ,
to achieve $100\%$ redshift completeness. For galaxies that have redshifts 
available from more than one sources, preferences are given in the order 
given above, i.e. from 2dFGRS to 2MPZ. As a result, $294$, $29$, $168$, $227$, 
$13,548$, and $1,985$ additional redshifts are obtained from 2dFGRS, 6dFGS, 
KIAS VAGC, LAMOST, the nearest neighbors, and 2MPZ, respectively. 
In the following, we will construct two different kinds of group catalogs from the 
SDSS data, one using only galaxies that have spectroscopic redshifts, 
and the other using all galaxies including the ones with estimated redshifts 
from nearest neighbors and from 2MPZ. For brevity, we refer to the 1st 
catalog as the SDSS and the 2nd as the SDSS+. We covert all the recession 
velocities (redshifts) to the CMB rest-frame, and restrict our samples to $z\leq 0.2$. 
This leaves a total of $586,025$ and $600,458$ galaxies as our final SDSS 
and SDSS+ samples, respectively. 

We compute the absolute magnitudes in the $r$-band of the sample according to 
equation (1), using the WMAP9 cosmology, with $K$- and evolution- corrections 
to $z=0.1$ following Poggianti (1997). We also calculate the $(g-r)$
color, corrected to $z=0.1$, for each galaxy. From the DR13 photometric catalog, 
we adopt the cmodel magnitude to calculate the flux, and the model magnitudes 
to compute the color, following the recommendations of the SDSS team. 
The zero-point offset between the DR13 magnitude and the AB magnitude is 
practically zero for $g$- and $r$- bands within the error of $0.01$ mag. 
Galaxies with colors outside the $3\sigma$ of the color distribution at 
a given luminosity are assigned the median color. For galaxies with 
negative redshifts, their luminosities are obtained from their EDD 
distances. Finally, we estimate the stellar masses of individual galaxies from 
their $r$-band absolute magnitudes and $(g-r)$ colors, following the formula 
of \citet{bell03}: 
%%%%%%%%%%%%%%%%%%%%%%%%%%%%%%%%%%
\begin{eqnarray}\label{eq_}
\log{M_*} = -0.306 + 1.097(g-r) + 0.4(4.67-M_r),  
\end{eqnarray}
%%%%%%%%%%%%%%%%%%%%%%%%%%%%%%%%%%%
where $4.67$ is the absolute magnitude of the Sun in the $r$-band.

\subsection{The 2dFGRS catalog} 
\label{ssec_2dFGRScat}
%%%%%HJM revised
%%%%%slim revised

Finally, we also select a sample from the 2dF Galaxy Redshift Survey 
\citep[2dFGRS;][]{colless01}. The 2dFGRS provides redshifts for about 
$250,000$ galaxies, measured with the Two-degree Field (2dF) multifibre 
spectrograph on the Anglo-Austrailian Telescope,
down to a magnitude limit of $b_J=19.45$ after Galactic extinction correction. 
The survey consists of two strips in the northern
and southern Galactic hemispheres (the northern and southern Galactic caps, 
respectively), and $99$ `random' fields of $2^\circ$ each over and around the 
southern Galactic cap. The full survey covers about $2,000{\rm deg^2}$ with 
a median redshift of $z\sim0.11$. Because the random fields are not
contiguous and our group finder can be affected severely at the edges
of these field, we use only the two Galactic caps for our purpose.  
The final sky coverage of our sample is about $3.5\%$. 

The quality of a spectrum is characterized by a quality parameter, $Q=1$ - $5$, 
with a higher value of $Q$ indicating higher quality. 
From the final release spectroscopic catalog, we use only galaxies with 
$Q\geq 3$, for which the redshifts are $98.4$ per cent reliable, with a 
typical uncertainty of $85\,{\rm km/s}$ \citep[][]{colless01}. 
Of all the $245,591$ galaxies from the 2dFGRS catalog, $12,340$ systems 
do not have spectroscopic redshifts from the survey. For these galaxies, 
we find matches and assign redshifts from the SDSS DR13, the 6dFGS, the 
nearest neighbors, and the 2MPZ. In the cases where a galaxy has redshifts 
from more than one of these sources, the priority is given, in the order 
of decreasing priority, to the SDSS, the 6dFGS, the nearest neighbor
redshift, and the 2MPZ. As a result, we have $233,251$ redshifts from 
the 2dFGRS, $322$ from the SDSS, $43$ from the 6dFGS, $11,852$ from 
the nearest neighbors, and $123$ from the 2MPZ. 
Again, we will provide two catalogs, 
one using only galaxies with spectroscopic redshifts, and 
the other using all galaxies. We refer to these two catalogs as 
the 2dFGRS and 2dFGRS+ samples, respectively. 

Redshifts are corrected to the CMB rest-frame, and we limit our sample 
to $z\leq0.2$. The final samples contain $180,967$ and $189,101$ galaxies 
for the 2dFGRS and 2dFGRS+, respectively. When selecting galaxy groups from 
these samples, we adopt the survey masks provided in the 2dFGRS website
\footnote{http://www.2dfgrs.net/}. For reference, the redshift distribution 
of 2dFGRS is shown in Figure~\ref{fig_z_dist}. 

We use equation (1) to convert the observed apparent magnitudes to 
absolute magnitudes, assuming WMAP9 cosmology. To do this, we first make $K$- 
and $E$- corrections to $z=0.1$ following the method given in \citet{poggianti97}.
The stellar masses of individual galaxies are obtained from 
their $b_J$-band absolute magnitudes and $b_J-R$ colors using the 
approximation of \citet{bell03}: 
%%%%%%%%%%%%%%%%%%%%%%%%%%%%%%%%%%
\begin{eqnarray}\label{eq_}
\log{M_*} = -0.976 + 1.111(b_J-R) + 0.4(5.48-M_{b_j}),  
\end{eqnarray}
%%%%%%%%%%%%%%%%%%%%%%%%%%%%%%%%%%%
where $5.48$ is the absolute magnitude of the Sun in the $b_J$-band.
For a small number of galaxies that have colors outside  the $3\sigma$ range 
of the $b_J-R$ distribution, and for a total of $276$ 
galaxies without the $R$-band photometry, each of them has been assigned a $(b_J-R)$ 
color that is equal to the median value given by the galaxies which have 
$b_J$ luminosities similar to the galaxy in question 
and have $b_J-R$ colors. Here again, the EDD distances have been used to convert
the observed flux to the luminosity for galaxies with negative redshifts.

\subsection[mock]{Mock samples used to test methods}
\label{ssec_mock}
%%%%HJM revised

The quality of the group samples to be constructed depends on 
the performance of the group finder used to identify the groups 
from the observational data. To test the performance of our group 
finder (to be described in \S\ref{sec_gfinder}), we use mock samples 
constructed from a hydrodynamical simulation of galaxy formation, 
where information about dark matter halos and their galaxy memberships 
are all known.

The hydrodynamical simulation used here is the Evolution and 
Assembly of GaLaxies and their Environments \citep[EAGLE;][]{schaye15, crain15, mcalpine15}. 
EAGLE follows the evolution of gas, stars, dark matter, and massive black holes
in a cosmological context, implementing physical models for gas 
cooling, star formation, stellar and AGN feedback. Sub-grid processes, 
in particular feedback processes, are modeled with simple parametric 
forms, with model parameters tuned to match observations,  
such as the stellar mass function and stellar mass - black hole 
mass relation at $z\sim0$, as detailed in \citet{crain15}.
The simulation starts from $z=127$ and adopts 
the {\it Planck} cosmology with $(\Omega_m, \Omega_\Lambda, h) = (0.307, 0.693, 0.678)$
\citep[][]{planck14}. This cosmological model is not 
exactly the same as the WMAP9 cosmology we adopt in this paper. 
However, since the purpose here is to test our group finder and 
halo mass proxies (see below), this difference in cosmology should not be a concern, 
as long as the analysis is done in a self-consistent way. 
EAGLE provides a set of simulations assuming different sets of 
model parameters and different box sizes. Here we use the simulation 
with the largest box size of $100{\rm Mpc^3}$, their fiducial simulation. 
The simulation contains about $11,500$ dark matter halos with masses above 
$10^{11} {\rm M}_\odot$, and $\sim 10,000$ galaxies with masses 
comparable to or above that of the Milky Way. EAGLE adopted
the \citet{chabrier03} IMF and the spectral synthesis model of 
\citet{bruzual03} to get luminosities and stellar masses of 
individual galaxies from 
their star formation histories, and these are used in our analysis as well.
A few galaxies are found to have extremely low halo masses for their 
stellar masses. These extreme outliers are excluded from our analysis. 

We construct realistic mock catalogs of galaxies from EAGLE. Since 
the original simulation box, $100 \,{\rm Mpc}$, 
is smaller than the volumes of our samples, we stack the duplicates of the 
original box side by side as many times as is required to 
cover the volume of the sample in question. A location 
is chosen for the observer in the stack, and apparent magnitudes of 
individual galaxies are calculated from their luminosities and their distances 
to the observer. The same sample selections 
as those for the observational samples, as detailed earlier in this section,
are applied to construct the mock samples. Specifically, we choose galaxies in the 
simulation box that are in the same sky regions as the observational samples, 
as well as apply the apparent magnitude and redshift limits of each survey 
to eliminate faint galaxies from the mock samples. Finally, we also apply the 
same masks, if any, as provided for the observational samples by each survey. 
Since all the galaxies in the simulation are linked to dark matter halos, we can 
use these mock catalogs to quantify the accuracy of our methods. 

As an independent check, we have also constructed mock samples using an 
empirical model of galaxy formation. The details of these mock samples 
are given in Appendix A.

%%%%%%% SECTION 3
\section[gfinder]{THE HALO-BASED GROUP FINDER}
\label{sec_gfinder}
%%%%%%%HJM revised

\subsection{The basic algorithm} 

The method adopted here is similar to the `halo-based' group finder   
developed by \citet{yang05} (Y05 hereafter). This group finder makes 
use of physical properties of dark matter halos expected from 
the current cold dark matter (CDM) cosmogony, such as halo mass, 
virial radius and velocity dispersion, in assigning galaxies into 
groups. The group finder has been tested extensively using mock galaxies,  
and is found to be more effective than the traditional Friends-of-Friends
(FoF) algorithm in grouping galaxies according to common halos, 
and particularly in dealing with poor groups associated with small halos.
This allows the identification of systems over a wide range of 
masses. However,  in the original group finder of Y05,  
halo masses assigned to galaxy groups are based on the 
ranking order of  the total luminosity of member galaxies 
(or the sum of the luminosities of member galaxies above a certain 
luminosity limit). It becomes inaccurate 
for groups that contain only a small number of members, and   
is not appropriate for shallow surveys where a large number 
of the identified groups contain only one or a small number of 
relatively bright galaxies. In order to overcome this limitation, 
we make some modifications to the group finder of
Y05, in particular in the assignments of masses to galaxy groups.
Specifically, for systems containing more than one member galaxy, 
we adopt a modified version of the `GAP' model developed by
\citet{lu16}. For systems containing only one member, we use 
halo mass proxies that are calibrated by realistic mock catalogs. As we will 
show below, these modifications not only provide more 
accurate halo mass estimates, but also allow us to reach 
to systems with lower halo masses in a uniform way.
The detailed steps of the group finder are as followings: 

\itemize{}
\item {Step 1.} {\it Assign preliminary halo mass to every galaxy.}
\newline

 While the group finder of Y05 starts by linking galaxies using the FoF 
algorithm with a small linking length to identify preliminary group centers, 
we start by treating all galaxies as isolated galaxies associated with distinct
tentative dark matter halos with preliminary halo masses computed according to the 
halo mass proxies described in \S\ref{ssec_massproxy}. 
We have checked that this leads to 
no significant differences in membership, mass, and the number of 
final groups in comparison to that of Y05. 
\newline

\item {Step 2.} {\it Membership assignment using halo properties.}
\newline

  For all groups identified at each iteration, we compute the size and the 
line-of-sight velocity dispersion, which are used to determine 
which galaxies should be assigned to a certain group,   
%%%%%%%%%%%%%%%%%%%%%%%%%%%%%%%%%%
\begin{eqnarray}
\frac{r_{180}}{{\rm Mpc}}\ \ \ \ &= &1.33\ h^{-1}\ \Big(\frac{M_h}
{10^{14}h^{-1}M_\odot}\Big)^{1/3} (1+z_{\rm group})^{-1} \nonumber 
\end{eqnarray}
%%%%%%%%%%%%%%%%%%%%%%%%%%%%%%%%%%%
%%%%%%%%%%%%%%%%%%%%%%%%%%%%%%%%%%
\begin{eqnarray}
\frac{\sigma}{\rm km\ s^{-1}}&= & 418\ \Big(\frac{M_h}
{10^{14}h^{-1}M_\odot}\Big)^{0.3367}\,,  
\end{eqnarray}
%%%%%%%%%%%%%%%%%%%%%%%%%%%%%%%%%%%
where $z_{\rm group}$ is the redshift of the group in question,
and $r_{180}$ is the radius of the halo, within which the mean 
mass density is 180 times the mean density of the universe at 
the given redshift. The numbers used are appropriate 
for the WMAP9 cosmology \citep[e.g.][]{lu16}. 
Next, we assume that the phase-space distribution of galaxies in dark matter 
halos follows that of dark matter particles and that the group center 
is the same as the halo center. The number density contrast of galaxies 
at the redshift of $z_{\rm group}$ can then be expressed as 
%%%%%%%%%%%%%%%%%%%%%%%%%%%%%%%%%%
\begin{eqnarray}
P_M(R,\Delta z) = \frac{H_0}{c}\frac{\Sigma(R)}{\bar{\rho}}p(\Delta z) 
\end{eqnarray}
%%%%%%%%%%%%%%%%%%%%%%%%%%%%%%%%%%%
where $R$ is the projected distance, $c$ is the speed of light, $\bar{\rho}$ is
the mean density of the Universe, $\Sigma(R)$ is the surface density, and $\Delta 
z=z-z_{\rm group}$. We assume that the redshift distribution of galaxies 
within a halo, $p(\Delta z)$, has the Gaussian form, 
%%%%%%%%%%%%%%%%%%%%%%%%%%%%%%%%%%
\begin{eqnarray}
p(\Delta z) = \frac{c}{\sqrt{2\pi}\sigma(1+z_{\rm group})}
\exp\Bigg(\frac{-c^2\Delta z^2}{2\sigma^2(1+z_{\rm group})^2}\Bigg) 
\end{eqnarray}
%%%%%%%%%%%%%%%%%%%%%%%%%%%%%%%%%%%
where $\sigma$ is the line-of-sight velocity dispersion. 
Furthermore, halos are assumed to follow a spherical NFW 
density profile, so that the surface density $\Sigma(R)$ can be written as  
%%%%%%%%%%%%%%%%%%%%%%%%%%%%%%%%%%
\begin{eqnarray}
\Sigma (R) = 2r_s\bar{\delta}\bar{\rho}f(R/r_s) 
\end{eqnarray}
%%%%%%%%%%%%%%%%%%%%%%%%%%%%%%%%%%%
where $r_s$ is the scale radius, and  
%%%%%%%%%%%%%%%%%%%%%%%%%%%%%%%%%%
\begin{eqnarray}
f(x) = 
\begin{cases} 
\frac{1}{x^2-1}\Bigg[1-\frac{\ln \frac{1+\sqrt{1-x^2}}{x}}{\sqrt{1-x^2}}\Bigg], 
& \mbox{if } x<1 \\ 
\frac{1}{3}, & \mbox{if } x=1 \\ 
\frac{1}{x^2-1}\Bigg[1-\frac{{\rm atan}\sqrt{x^2-1}}{\sqrt{x^2-1}}\Bigg], 
& \mbox{if } x>1 \nonumber
\end{cases}
\end{eqnarray}
%%%%%%%%%%%%%%%%%%%%%%%%%%%%%%%%%%%
%%%%%%%%%%%%%%%%%%%%%%%%%%%%%%%%%%
\begin{eqnarray}
\bar{\delta} = \frac{180}{3}\frac{c_{180}^3}{\ln (1+c_{180})-c_{180}/(1+c_{180})} 
\end{eqnarray}
%%%%%%%%%%%%%%%%%%%%%%%%%%%%%%%%%%%
with the concentration, $c_{180}=r_{180}/r_s$, given by the model of  
\citet{zhao09}. Finally,  we calculate $P_M(R, \Delta z)$ for each 
of all the galaxy-group pairs. If the value of $P_M$ is above a certain 
background value, $P_B$, an association between the galaxy and the 
group is assumed. If a galaxy is associated  with more than one group 
according to this criterion, the galaxy is assigned to the group 
with the largest $P_M(R, \Delta z)$. As demonstrated in Y05 using 
realistic mock samples, a compromise between the completeness and contamination 
can be achieved with $P_B\sim 10$, and the performance of the group finder 
is not very sensitive to the exact value of $P_B$. Note that $P_B\sim10$ 
is also in agreement with theoretical expectations for dark matter halos 
(see the discussion in section 3.2 of Y05 for details). 
We therefore adopt $P_B=10$ throughout this paper. 

After the membership of a group is determined, we define the 
stellar mass-weighted center of member galaxies as the group center, 
if stellar masses are available. Otherwise, we use luminosity-weighted 
center as the group center. 
\newline

%%%% Table ?? %%%%
\begin{table*}
 \renewcommand{\arraystretch}{1.5} 
 \centering
 \begin{minipage}{97mm}
  \caption{The best-fit parameters for the GAP correction from the EAGLE.}
  \begin{tabular}{cccccccc}
\hline
Sample & $\alpha_1$ & $\alpha_2$ & $\beta$ & $\gamma$ & $\delta_1$ & $\delta_2$ & 
$\delta_3$ \\
 & $(\times 10^5)$ & $(\times 10^6)$ & & & & &  \\
\hline
\hline
2MRS & $-2.3$ & $-1.1$ & $6.6$ & $5.1$ & $0.28$ & $0.13$ & $0.035$ \\

6dFGS & $-1.1$ & $-3.9$ & $4.8$ & $6.1$ & $0.13$ & $-0.15$ & $0.037$ \\

SDSS & $-3.6$ & $-6.0$ & $5.3$ & $5.2$ & $0.11$ & $-0.039$ & $0.0014$ \\

2dFGRS & $-3.6$ & $-5.9$ & $5.5$ & $5.7$ & $0.12$ & $-0.046$ & $0.0035$ \\

\hline
%\vspace{-2mm}
\end{tabular}
%\textbf{Notes.}
\vspace{3mm}
\label{tab_params}
\end{minipage}
\end{table*}
%%%%%%%%%%%%%%%%%%%%%%%%%%%%%%%%%%%%%%%%%%%%%%%%%%%%%%%%%%%%%%%%%%%%%%%%%%%%%%

\item {Step 3.} {\it Rank groups according to halo mass proxies.}
\newline

 As described in \S\ref{ssec_proxyforrich} and \S\ref{ssec_proxyforpoor}, 
in the beginning of each iteration, tentative halo masses are assigned to groups
identified in the previous step by ranking groups according to 
a mass proxy.  In short, for tentative groups containing only 
one member galaxy in the previous step, we use the galaxy stellar mass 
(luminosity) - halo mass relation obtained from a hydrodynamical 
simulation to assign the preliminary halo mass, as described in 
\S\ref{ssec_proxyforpoor}. For tentative groups that contain more than 
one member at a given iteration, we use the `GAP correction' method of 
\citet{lu16},  modified with our own re-calibrations (see below).  
\newline

\item {Step 4.} {\it Group mass update and iteration.}
\newline

 To assign masses to groups, we use abundance matching between 
the mass function of the preliminary groups and an adopted  
theoretical halo mass function. A new halo mass, $M_{\rm halo}$, 
is assigned to a group to replace the preliminary group mass, 
$M_{\rm pre}$, according to  
%%%%%%%%%%%%%%%%%%%%%%%%%%%%%%%%%%
\begin{eqnarray}
N(>M_{\rm halo}) = N(>M_{\rm pre}) \nonumber
\end{eqnarray}
%%%%%%%%%%%%%%%%%%%%%%%%%%%%%%%%%%%
where $N$ is the cumulative number density of groups (halos) more massive 
than $M_{\rm pre}$ ($M_{\rm halo}$). We use the theoretical halo mass function 
of \citet{sheth01} for this. Note that for flux-limted samples, 
halos of a given mass are complete only to a certain redshift. 
The abundance matching used to assign halo mass is applied only for groups 
in  samples that are complete (see \S\ref{ssec_mass}). For groups residing  
in volumes within which the samples are not complete, 
we use the mean relation between the halo mass and the mass proxy from 
the last iteration to assign halo masses to them. 
Once group masses are updated, 
we iterate Steps 2 through 4 until convergence in group membership is 
achieved. 

\subsection{Halo mass proxies of galaxy groups}
\label{ssec_massproxy}

As mentioned in the previous section, our group finder relies on 
the reliability of the halo mass model for groups. Here we test 
different halo mass proxies by comparing their predictions with 
the results obtained from the hydrodynamical simulation. We have also made 
similar tests using a mock sample of galaxies constructed by 
applying the empirical model of \citet{lu15} to the simulated 
halos. The results obtained from the empirical model are very similar 
to those obtained from the hydrodynamical simulation, and are presented 
in Appendix~\ref{sec_appendix}.

\subsubsection{Halo mass proxies for groups containing more than one galaxy}
\label{ssec_proxyforrich}

%%%%%%%%%%% figure3
\begin{figure*}
\includegraphics[width=0.9\linewidth]{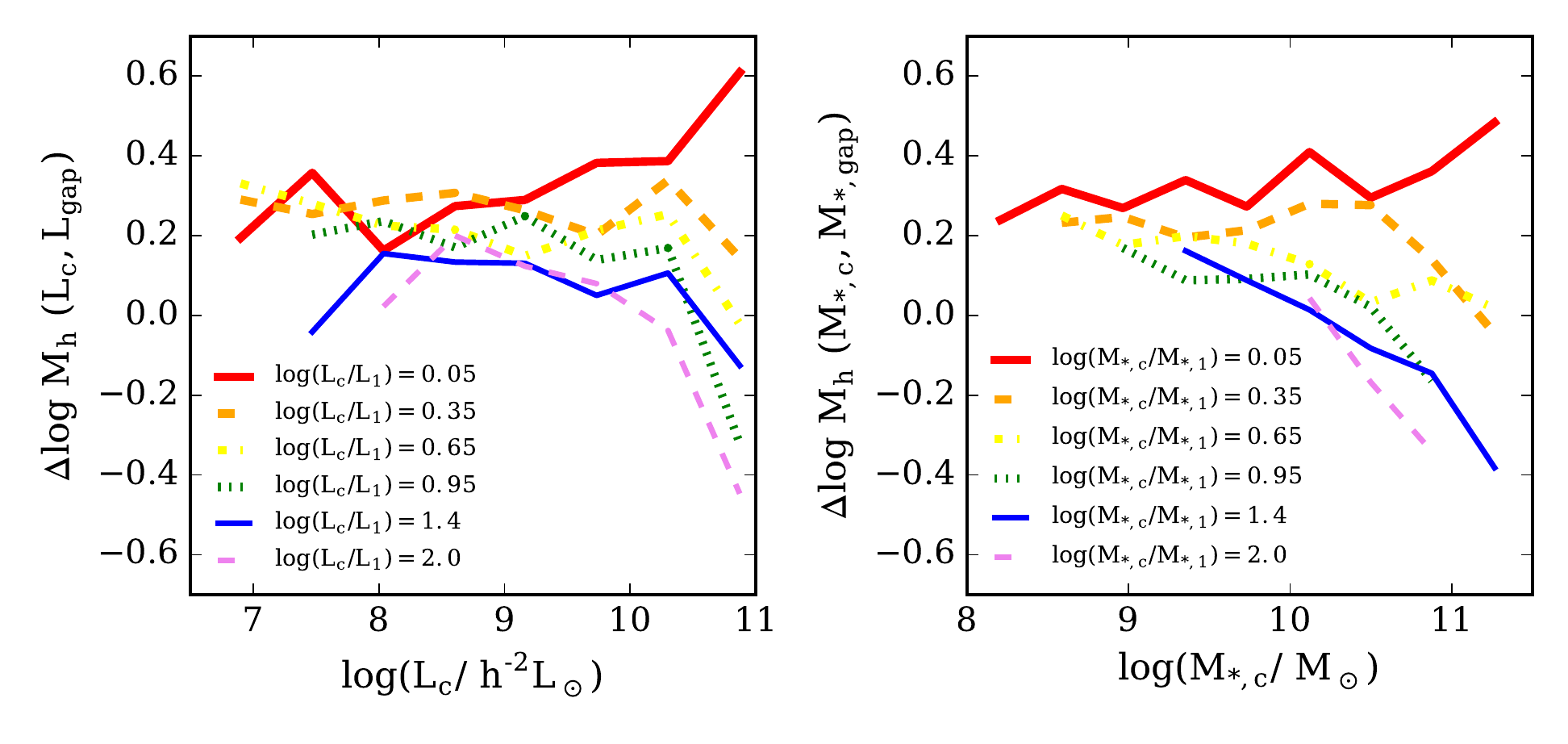}
\caption{Dependence of halo mass on the luminosity or stellar mass of 
the central galaxy and the GAP parameter (defined to be 
the difference in luminosity or stellar mass between the 
central galaxy and the $n$-th brightest satellite; 
see text for detailed definition) as given by EAGLE.
The left panel is the result based on the $K$-band luminosity, 
while the right panel shows result based on stellar mass. 
For clarity, only the GAP correction using the brightest satellite
(i.e. $n=1$) is shown. }
\label{fig_GAP}
\end{figure*}

In the original paper presenting the halo-based group finder, Y05 
uses the sum of the luminosities of member galaxies down to some luminosity 
limit as a proxy of group masses. However, this proxy may not be 
appropriate for a shallow galaxy survey where many groups 
have only a small number of members. Because of this, \citet{lu16}
(L16 hereafter) suggested the use of a combination 
of the luminosity/stellar mass of the central galaxies, and the 
luminosity/stellar mass GAP (the difference
in luminosity/stellar mass between the central galaxy and the $n$-th 
brightest galaxy) as a group mass proxy. Based on mock galaxy samples,
L16 came up with the following model for the halo mass, 
%%%%%%%%%%%%%%%%%%%%%%%%%%%%%%%%%%
\begin{eqnarray}
\log M_h(L_c, L_{\rm gap}) = \log M_h(L_c) + \Delta \log M_h(L_c, L_{\rm gap}) 
\end{eqnarray}
%%%%%%%%%%%%%%%%%%%%%%%%%%%%%%%%%%%
where $L_c$ is the luminosity of the central galaxy, $M_h(L_c)$ 
is the mean halo mass at a given $L_c$, and $L_{\rm gap}=L_c/L_{n}$ 
with $L_n$ the luminosity of the $n$-th brightest satellite.
The masses and luminosities here are in units of ${\rm M_\odot}/h$ and 
${\rm L_\odot}/h^2$, respectively.
Using mock samples constructed for the 2MRS, L16 found that their group 
masses are consistent with true halo masses obtained from 
the simulation used in their mock samples, and the best 
result is achieved with $n=4$. For groups with less than four 
satellites, L16 used the faintest satellite in a group for the GAP 
correction. The basic motivation behind the GAP correction is that 
groups of the same central galaxy luminosity but with more 
contribution from satellites should possess more massive halos. 

Here we adopt the same idea of the GAP correction, but 
use our own functional form for $\Delta \log M_h(L_c, L_{\rm gap})$:
%%%%%%%%%%%%%%%%%%%%%%%%%%%%%%%%%%
\begin{eqnarray}
\Delta \log M_h(L_c, L_{\rm gap}) = \alpha(L_{\rm gap}) \times (\log 
L_c - \beta)^\gamma + \delta(L_{\rm gap}) \nonumber
\end{eqnarray}
%%%%%%%%%%%%%%%%%%%%%%%%%%%%%%%%%%%
with 
%%%%%%%%%%%%%%%%%%%%%%%%%%%%%%%%%%
\begin{eqnarray}
\alpha(L_{\rm gap}) &=& \alpha_1 + \alpha_2 \log(L_{\rm gap}) \nonumber \\
\delta(L_{\rm gap}) &=& \delta_1 + \delta_2 \log(L_{\rm gap}) + 
\delta_3 [\log(L_{\rm gap})]^2
\end{eqnarray}
%%%%%%%%%%%%%%%%%%%%%%%%%%%%%%%%%%%
where the free parameters $\alpha_1$, $\alpha_2$, $\beta$, $\gamma$, 
$\delta_1$, $\delta_2$, and $\delta_3$ are constants, 
and the masses and luminosities are again in solar units. 
We use mock catalogs constructed from EAGLE to calibrate these 
free parameters. For example, from the 2MRS mock catalog 
we obtain $(\alpha_1, \alpha_2, \beta, \gamma, \delta_1, \delta_2, \delta_3)=
(-2.3\times10^{-5},\ -1.1\times10^{-6},\ 6.6,\ 5.1,\ 0.28,\ 0.13,\ 0.035)$ 
for $L_{\rm gap}=L_c/L_{2}$. The values of these parameters for other cases
are given in Table~\ref{tab_params}. 

As shown later in this section, the use of stellar masses 
gives better halo proxies than the use of luminosities.  
Thus, halo masses based on stellar masses are preferred to 
those based on luminosities whenever stellar masses are available. 
The halo mass proxy using stellar mass 
is modeled in the same way as that given above, except with $L_c$ and 
$L_{\rm gap}$ replaced by $M_{*,c}$ and $M_{\rm *, gap}$, respectively. 
Figure~\ref{fig_GAP} shows the relations given by equations (9) 
and (10) using galaxy luminosities or galaxy stellar masses. 

\subsubsection{Halo mass proxies for groups containing one galaxy}
\label{ssec_proxyforpoor}

Next we consider systems that contain only one member galaxy.  
Here we present the best proxy for such systems for each catalog
based on tests with a number of proxies. Note that the GAP correction 
from the previous section is not applicable for isolated galaxies, 
as by definition there is no observed satellite in systems 
containing only one member. In L16, it was assumed that each 
isolated galaxy has, with $50\%$ chance, one potential satellite 
galaxy with $K_s=11.75$, which is the magnitude limit of 2MRS catalog.  
The average of the corresponding GAP-corrected group mass 
and $M_h(L_c)$ were used as the halo mass proxy in L16, if 
the GAP-correction $\Delta\log M_h$ is larger than $0.5$. 
Such a prescription sometimes leads to too high or too low a 
halo mass for a given $L_c$ according to our test with the mock 
samples used here. Because of this, here we attempt to revise the 
proxy so that it is more reliable for groups with only one member. 

%%%%%%%%%%% figure4
\begin{figure*}
\includegraphics[width=0.75\linewidth]{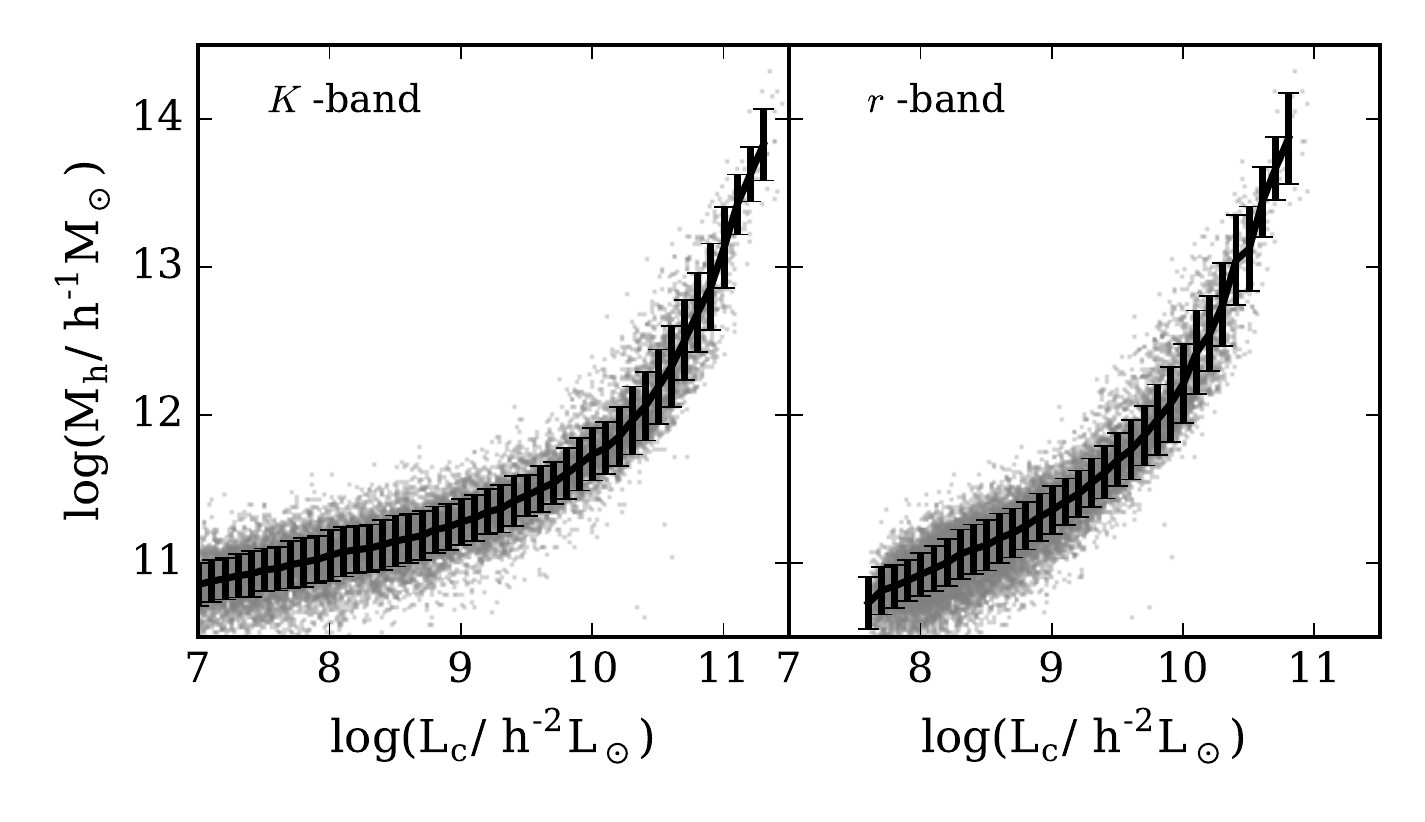}
\caption{The relation between the $K$-band (left) and the $r$-band (right) 
luminosity of central galaxies and the halo mass, as obtained from the 
EAGLE simulation. The gray points are individual systems, while the black 
line and bars show the median and scatter of the relation, respectively.}
\label{fig_Lc_Mh}
\end{figure*}

%%%%%%%%%%% figure5
\begin{figure*}
\includegraphics[width=0.75\linewidth]{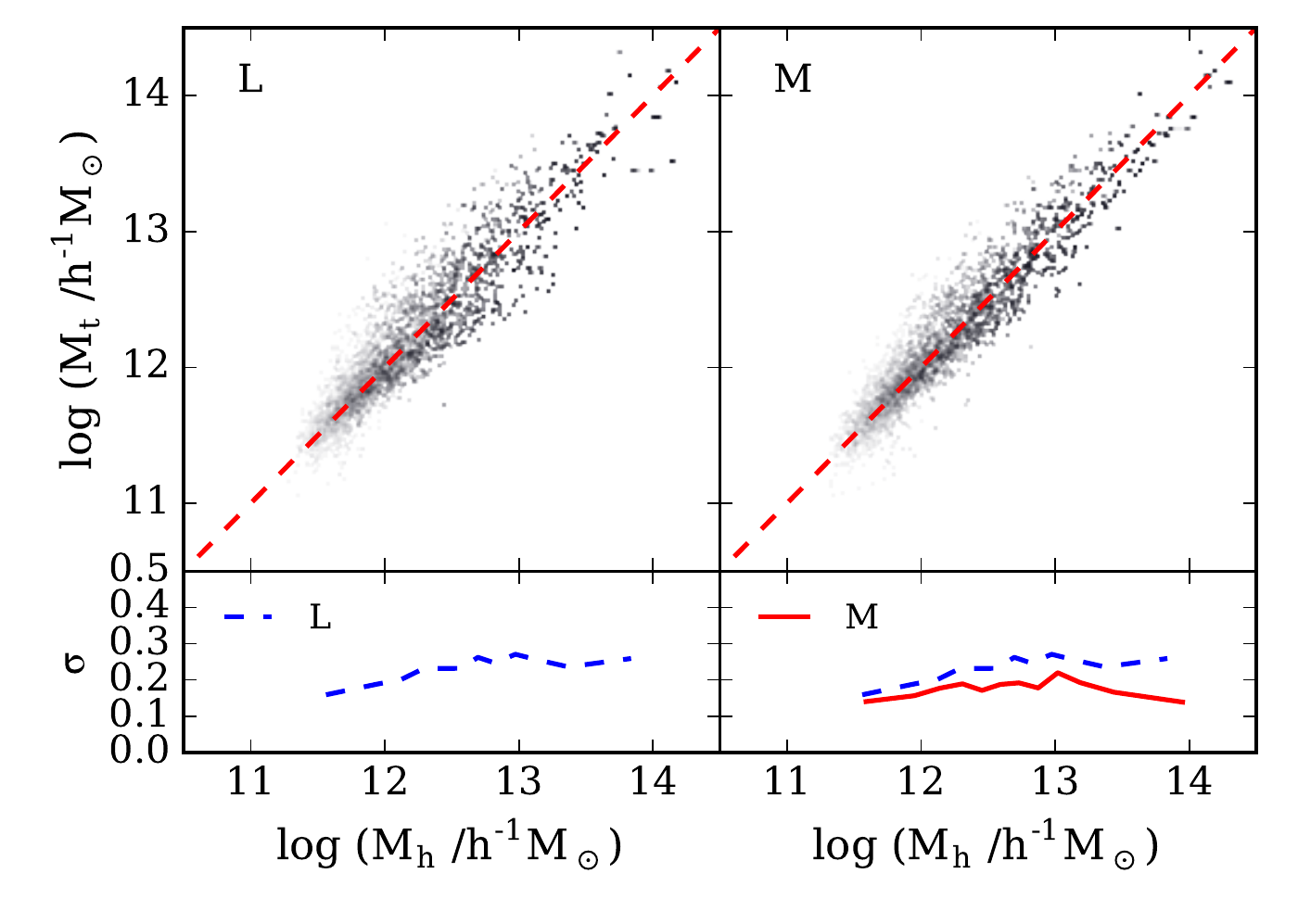}
\caption{The correlations of the halo masses given
by a mass proxy for groups containing a single member galaxy 
(horizontal axis) with the true halo mass (vertical axis), 
obtained from the 2MRS mock sample constructed with the EAGLE 
simulation. The results shown use proxies based on 
the $K$-band luminosity (L; left), and stellar mass (M; right). 
The red straight line in each big panel shows 
a perfect correlation, while the curves in the smaller 
panels show the scatter in the correlation.}
\label{fig_proxy}
\end{figure*}

%%%%%%%%%%% figure6
\begin{figure*}
\includegraphics[width=0.75\linewidth]{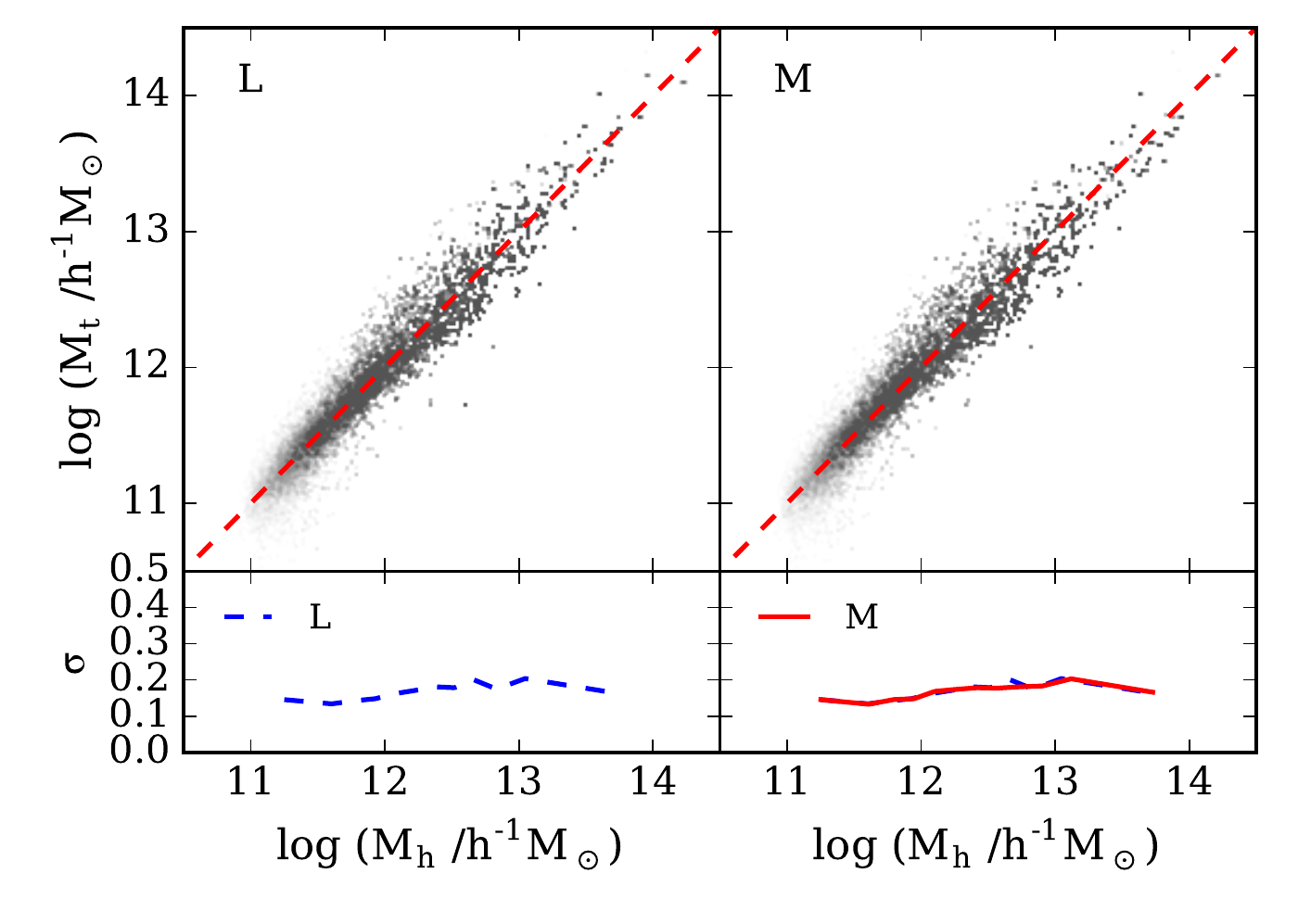}
\caption{The same comparison as in Figure~\ref{fig_proxy} but 
here for the SDSS mock sample.}
\label{fig_proxy_SDSS}
\end{figure*}

\paragraph{Proxy-L: Galaxy luminosity}

Our first halo mass proxy is based on the luminosities of 
galaxies. To do this, we first obtain the luminosity - halo mass 
relation of central galaxies from EAGLE. 
Figure~\ref{fig_Lc_Mh} shows such relations in the $K$-band and $r$-band. 
The distribution of the halo masses at a given luminosity 
is roughly log-normal. In the $K$-band, which will be 
used for both 2MRS and 6dFGS, the mean relation can be 
well described by 
%%%%%%%%%%%%
\begin{equation}
\log M_h
= 10.789 + 2.109\times10^{-4}\exp{(\log L_c/1.184)}
\end{equation}
%%%%%%%%%%%%%
and the typical width is about $0.2\,{\rm dex}$. The units of $M_h$ and
$L_c$ are in ${\rm M_\odot}/h$ and ${\rm L_\odot}/h^2$, respectively. 
In the $r$-band, which will be used for SDSS and 2dFGRS, the mean
relation is given by  
%%%%%%%%%%%%
\begin{equation}
\log M_h
= 10.595 + 4.370\times10^{-4}\exp{(\log L_c/1.214)}
\end{equation}
%%%%%%%%%%%%%
and the width of the log-normal distribution is about $0.22\,{\rm dex}$. 
We assign the mean halo mass at a given luminosity as the tentative 
halo mass to each galaxy. We also tested generating a random
mass at given luminosity around the mean halo mass and using it
as the mass proxy, and found that the resulting scatter 
between the true halo mass and final halo mass from the group finder 
is larger by $\sim0.1\,{\rm dex}$ than that given by using the mean relation. 

Figure~\ref{fig_proxy} compares the group masses given by 
the group finder with true halo mass from EAGLE for the 2MRS mock sample. 
The overall agreement between true mass and group mass
using this proxy is found to have scatter of $0.2-0.25\,{\rm dex}$. 
Note that Figure~\ref{fig_proxy} only includes isolated 
galaxies, which are expected to have larger scatter than groups of more than 
one member for which the GAP correction will be used. 

\paragraph{Proxy-M: Galaxy stellar mass}

We also test a halo mass proxy based on stellar mass of the central galaxy. 
To do this, we use the mean relation between the halo mass and stellar 
mass of isolated galaxies from EAGLE to assign preliminary halo mass. 
Figure~\ref{fig_proxy} shows the comparison of the resulting final group 
mass with true halo mass. It is clear that the scatter in the group mass 
is significantly reduced, by $\sim0.05\,{\rm dex}$ or more, relative to Proxy-L, 
suggesting that stellar mass is a better halo mass proxy for isolated 
galaxies. In real observations, however, stellar mass 
estimates introduce additional uncertainties. Thus we provide catalogs
based on both Proxy-L and Proxy-M in \S\ref{sec_gcatalog}, where we
construct and present our group catalogs. 

Figure~\ref{fig_proxy_SDSS} shows the same comparisons between the true and 
estimated halo masses for the SDSS mock sample. The two mass proxies, 
Proxy-L and Proxy-M are used in the same way as described above for the 2MRS 
mock, except that the parameters in the mass models are obtained for the 
SDSS $r$ magnitude. We see that for the SDSS mock catalog, the two 
mass proxies give very similar scatter in the halo mass, 
$\sim0.15-0.2\,{\rm dex}$. This is different from 2MRS, for which 
Proxy-M appears to be significantly more accurate than Proxy-L. 
This may be due to the fact that isolated galaxies in the SDSS mock 
are dominated by low-mass galaxies (because of its fainter magnitude limit)
for which the galaxy color does not depend systematically on halo mass.
We also found the same level of scatter, 
with perhaps a slight increase at the massive end, in tests based on the mock 
samples constructed from the empirical model, where uncertainties 
such as that in the stellar mass measurements, are taken into account. 

\paragraph{Other proxies tested}

We have tested a number of other quantities available from the EAGLE such as
velocity dispersion and metallicity of galaxies, as well as halo formation 
time and local density of galaxies. While some of these quantities are also 
found to be strongly correlated with halo mass according to the simulations, 
we found that halo mass proxies based on these quantities are not as 
accurate as those given by stellar mass and luminosity. We have also 
tested using combinations of stellar mass/luminosity and one of these 
additional quantities as halo mass proxies and found that none of them makes
significant improvement in the halo mass estimate, at least according to 
the simulation we use here. Note also that these additional quantities 
are usually not available from actual observations, making them less 
useful in practice.  

Motivated by L16 who used a `GAP limit' as a second parameter in the halo 
mass proxy for isolated galaxies,  we have also made tests with the use 
of some measurements of the `GAP limit'. The GAP limit, as defined in L16,  
is the GAP correction described in the previous section but using 
$G_{\rm gap, lim}=L_c/L_{\rm lim}$ instead of $G_{\rm gap}=L_c/L_{n}$, 
where $L_{\rm lim}$ is the luminosity that corresponds to the 
observational magnitude limit at the redshift of the galaxy in question.
Thus, an isolated galaxy with smaller $G_{\rm gap, lim}$ should have, 
on average, a more massive satellite that is not observed due to the 
magnitude limit.
The `GAP limit' is an attempt to take such an effect into account. 
However, our test showed that using `GAP limit' does not lead to 
further improvement in the final halo mass. 

Given all these test results, we use the stellar mass when it 
is available, and use luminosity otherwise, as the halo mass 
proxy for isolated galaxies. However, given the observational uncertainties
in stellar mass estimates. we will provide two catalogs for each data set: 
a catalog constructed based on Proxy-L and a catalog based on Proxy-M.

%%%%%%% SECTION 4
\section[test]{TESTING THE GROUP FINDER WITH MOCK SAMPLES}
\label{sec_test}

%%%%%%%%%%% figure7
\begin{figure*}
\includegraphics[width=0.9\linewidth]{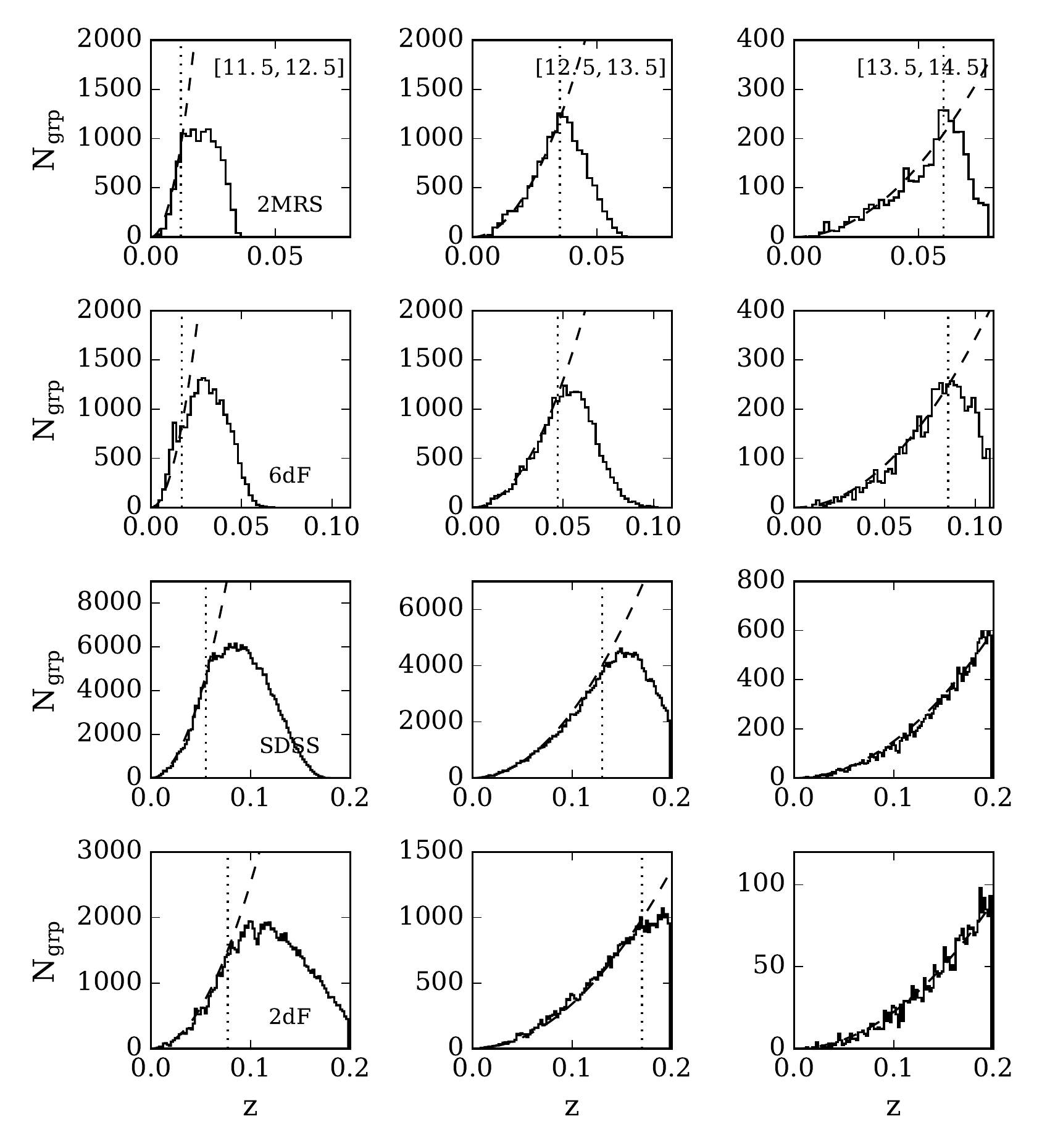}
\caption{The number of halos as a function of redshift from the EAGLE 
simulation in several mass bins as indicated in the upper panels for 
each mock sample. The vertical dotted lines show the redshift limits 
to which the samples are complete for a given halo mass. }
\label{fig_mock_Mcomp_z_how}
\end{figure*}

Before we apply the group finder to observational samples, 
we test its performances by applying it to realistic mock samples 
described in \S\ref{ssec_mock}, and analyzing the accuracy of 
group masses, and the completeness, contamination and purity of group 
memberships that are expected from each of the observational samples.   

%%%% Table ?? %%%%
\begin{table*}
 \renewcommand{\arraystretch}{1.5} 
 \centering
 \begin{minipage}{142mm}
  \caption{A summary of group catalogs constructed from the mock samples.\textsuperscript{a}}
  \begin{tabular}{cccccc}
\hline
Mock & Total groups & 
$N\textsuperscript{b}=1$ 
& $N=2$ & $N=3$ & $N\geq 4$ \\
\hline
\hline
2MRS & $30,124\ (29,462)$ & $25,160\ (24,305)$ & $3,128\ (3,310)$ & 
$881\ (845)$ & $955\ (1002)$ \\

6dFGS & $49,796\ (48,496)$ & $41,623\ (39,938)$ & $5,119\ (5,427)$ & 
$1,505\ (1,477)$ & $1,549\ (1,654)$ \\

SDSS & $473,303\ (454,474)$ & $397,300\ (377,280)$ & $49,166\ (51,832)$ & 
$13,937\ (14,587)$ & $14,846\ (15,847)$ \\

2dFGRS & $163,413\ (156,825)$ & $134,996\ (126,670)$ & $17,960\ (18,548)$ & 
$5,108\ (5,740)$ & $5,349\ (5,867)$ \\

\hline
\vspace{-2mm}
\end{tabular}
\textbf{Notes.}

a. The numbers in parentheses are from the simulation. 

b. The number of member galaxies in a group.
\vspace{-5mm}
\label{tab_mock}
\end{minipage}
\end{table*}
%%%%%%%%%%%%%%%%%%%%%%%%%%%%%%%%%%%%%%%%%%%%%%%%%%%%%%%%%%%%%%%%%%%%%%%%%%%%%%

\subsection{Applying the group finder to the mock samples}

As we have shown, stellar mass is theoretically a better 
proxy of group mass than luminosity (see \S\ref{sec_gfinder}) but only when 
observational uncertainties in stellar mass estimate are negligible. 
Therefore, we present catalogs that use both luminosity and stellar mass 
as mass proxies. We use the GAP correction in luminosity (or stellar mass) for 
groups with more than one member, and Proxy-L (or Proxy-M) for isolated galaxies. 

\subsection{Group mass estimates}
\label{ssec_mass}
%%%HJM revised

%%%%%%%%%%% figure8
\begin{figure}
\includegraphics[width=0.9\linewidth]{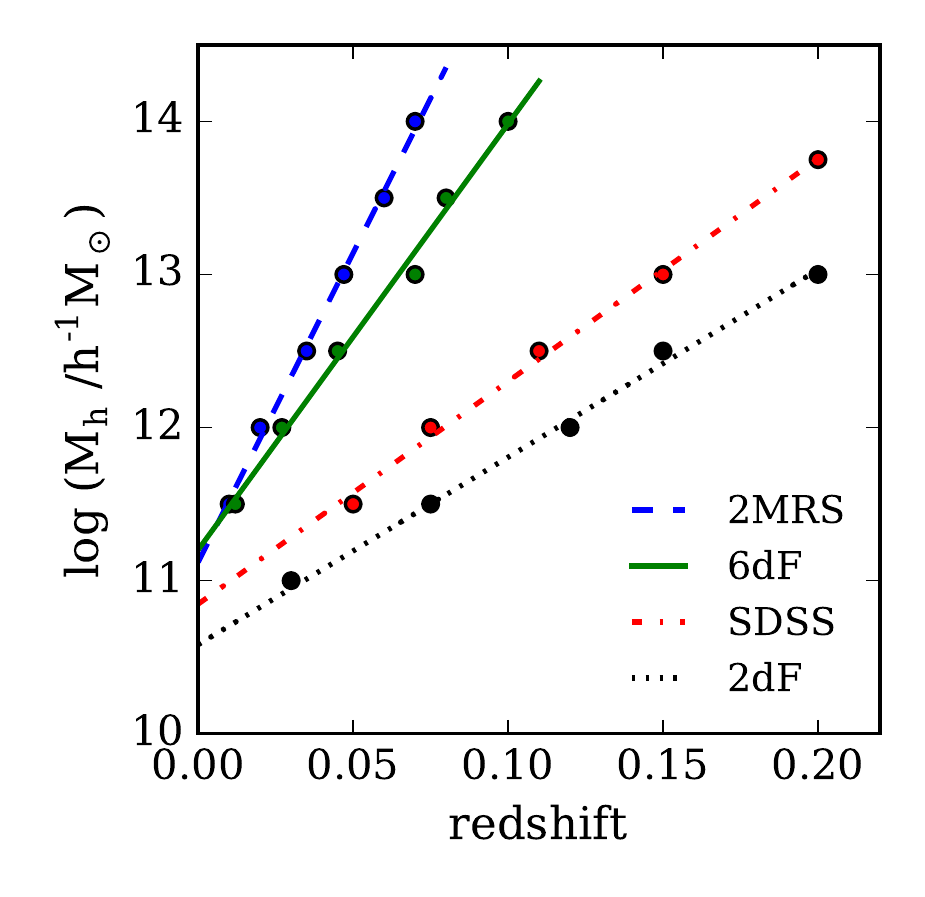}
\caption{Halo mass that is complete as a function of redshift for the mock 
samples of the simulation (circles), and linear fits to it (lines). We use
the linear relation for abundance matching to assign halo masses to the mock
groups. }
\label{fig_mock_Mcomp_z}
\end{figure}

%%%%%%%%%%% figure9
\begin{figure*}
\includegraphics[width=0.75\linewidth]{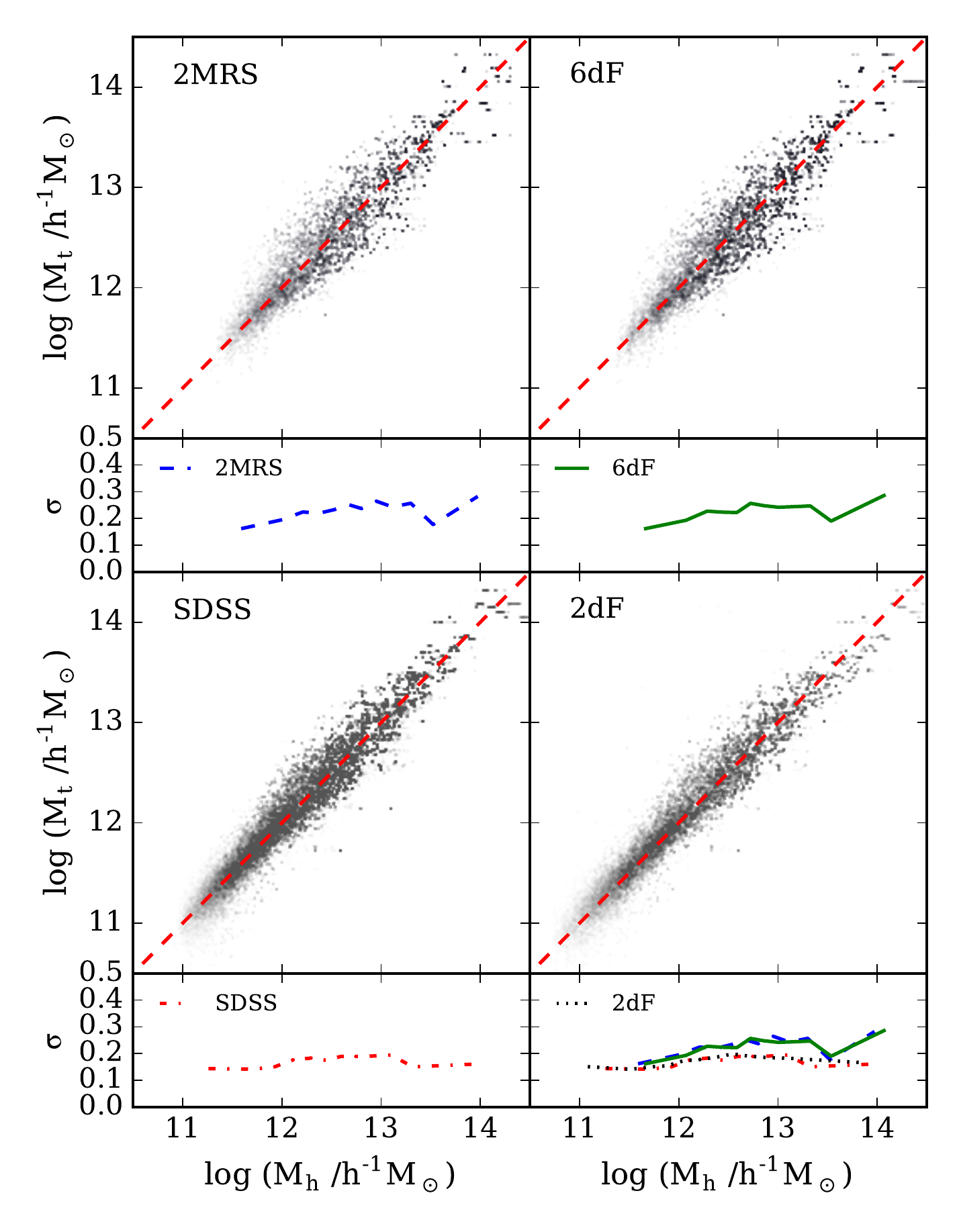}
\caption{Comparison between true halo mass (vertical
axis) and group mass identified by our group finder (horizontal axis) using
luminosity as the proxy of halo mass for
the mock samples of 2MRS, 6dFGS, SDSS, and 2dFGRS constructed 
from the EAGLE simulation (see text for the sample selections). The small
rectangular panels plot the scatter of true halo mass at given group mass. }
\label{fig_mock_mass_L}
\end{figure*}

%%%%%%%%%%% figure10
\begin{figure*}
\includegraphics[width=0.75\linewidth]{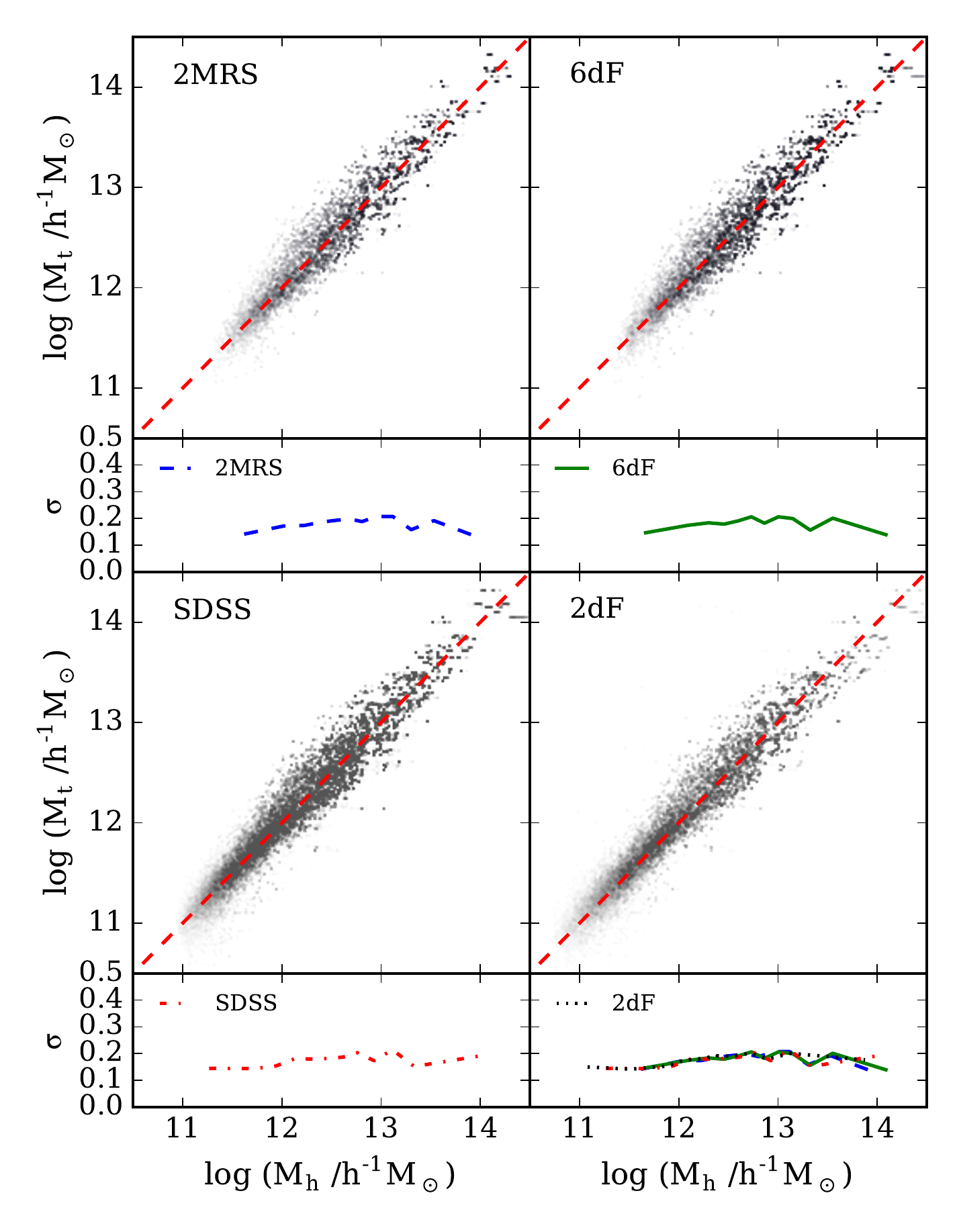}
\caption{Same comparison as Figure~\ref{fig_mock_mass_L} but using stellar 
mass as the proxy of halo mass.}
\label{fig_mock_mass_M}
\end{figure*}

%%%%%%%%%%% figure11
\begin{figure*}
\includegraphics[width=0.9\linewidth]{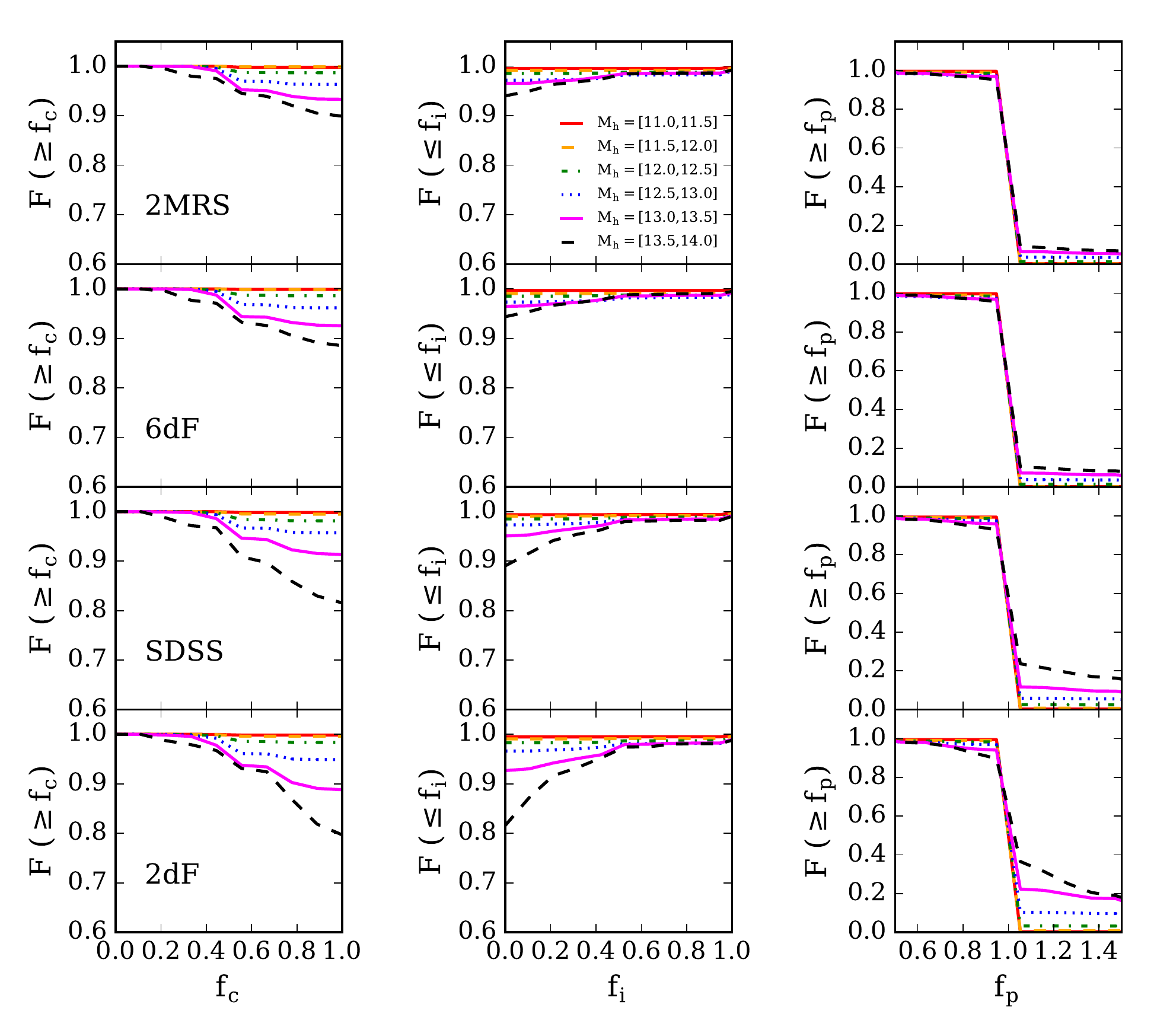}
\caption{Membership assignments by the group finder applied  to 
the mock samples, in terms of the completeness (left), contamination (middle), 
and purity (right). 
The vertical axis plots the cumulative fraction
of the groups identified via the group finder, and the different lines are for 
halos of different masses as indicated. }
\label{fig_mock_ccp}
\end{figure*}

%%%%%%%%%%% figure12
\begin{figure*}
\includegraphics[width=0.8\linewidth]{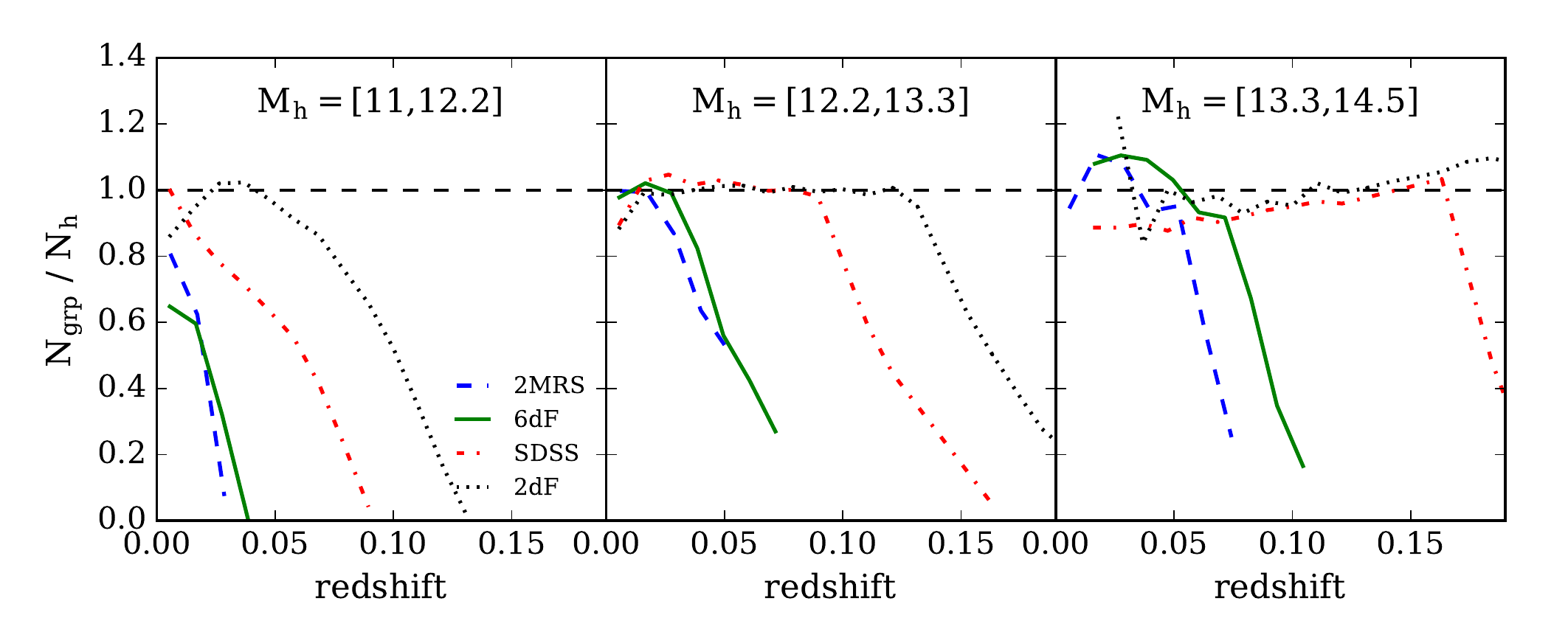}
\caption{Global completeness, defined as the number of groups identified
relative to the number of true halos, for the mock samples as a function of 
redshift for halos of different masses as indicated. }
\label{fig_mock_comp}
\end{figure*}

In the end of each iteration of the group finder, we finalize group masses
using abundance matching. The abundance matching is applied only to volumes within 
which groups of a given mass are complete. As the surveys and the mock samples are 
flux-limited, halos of a given mass are only complete to a certain redshift. 
Figure~\ref{fig_mock_Mcomp_z_how} shows the number of halos as a function
of redshift for each mock sample from the simulation. As one can see, 
in each case the number of groups first follows well the expectation of a 
constant density indicated by the dashed curve in each panel, and starts 
to go below the expectation at some redshift as incompleteness becomes 
severe. We can therefore define a limiting redshift, within which the
group sample in question is approximately complete. The limiting redshift, 
$z_{\rm lim}$, is indicated as the dot vertical line in each panel, and  
Figure~\ref{fig_mock_Mcomp_z} shows the value of $z_{\rm lim}$ as a 
function of halo/group mass for the four mock samples corresponding 
to the four observational samples. These relations can all be well 
described by a power law, $(1+z_{\rm lim})\propto M_{\rm h}^\zeta$,    
as shown in Figure~\ref{fig_mock_Mcomp_z}. The
limiting redshifts obtained in this way are used to define complete 
samples for abundance matching. For groups that 
are outside the limiting redshift, we use the mean 
relation between halo mass and the mass proxy to assign halo masses to them.  

Figure~\ref{fig_mock_mass_L} compares the true halo masses from the simulation 
with the final group masses obtained by our group finder using the $K_s$-band
($r$-band, for the SDSS and 2dFGRS) luminosity as the mass proxy. 
It is clear that the group finder performs quite well in assigning 
correct masses to groups over the whole range of halo mass for
various samples. No significant bias is seen in the assigned mass for 
any of the samples. The horizontal stretching of the data points 
appearing at the massive end is due to the stacking of the simulation 
box and the small number of massive halos in the original simulation  
box. The true halo masses are exactly the same for some of the halos 
that are the duplicates of the same halo in the original simulation, 
but the group masses assigned to them can be different because they 
are located at different redshifts in the mock sample. 
For a given true mass, the typical scatter in the assigned mass 
is $\sim0.2\,{\rm dex}$. The scatter is larger for the 2MRS and 6dFGS mock
samples, reflecting the less tight $K_s$-luminosity vs. halo-mass 
relation than the $r$-luminosity vs. halo-mass relation in the simulation.
Table~\ref{tab_mock} compares the total number of groups (halos) and 
the number of groups (halos) of given richness between the mock 
group catalogs and the original simulation.

Figure~\ref{fig_mock_mass_M} shows the same comparison of halo 
mass but obtained using stellar mass as the mass proxy for all surveys.
It is seen that, unlike in Figure~\ref{fig_mock_mass_L}, the scatter
in halo mass is almost identical for all surveys, and that stellar mass
performs as a better mass proxy than the $K_s$-band luminosity by $\sim0.05\,{\rm dex}$ 
or more, as seen earlier in Figure~\ref{fig_proxy}, while the $r$-band 
luminosity is an as good proxy as the stellar mass for the deeper surveys. 

\subsection{Completeness, contamination, and purity}
\label{ssec_purity}
%%%%% slim revised
%%%%% HJM revised

In addition to group masses, comparisons are also made between the 
membership assignment by the group finder and the true membership 
given by the simulation. To do this, we first assume that each mock 
group identified corresponds to the simulation halo that is 
associated with the brightest member of the mock group. As a 
quantitative assessment of the membership assignment, 
we follow \citet{yang07} and define the following quantities,
\begin{itemize}
    \item Completeness: $f_c \equiv N_s/N_t$;
    \item Contamination: $f_i \equiv N_i/N_t$;
    \item Purity: $f_p \equiv N_t/N_g$
\end{itemize}
where $N_t$ is the total number of member galaxies of each halo from 
the simulation, $N_s$ is the number of member galaxies of the corresponding
mock group that are true members of the simulation halo (thus $N_s\leq N_t$), 
$N_i$ is the number of member galaxies of the mock group that are not true
members of the simulation halo, and $N_g=N_i+N_s$ is the 
total number of members of the mock group. For a perfect group finder, 
$N_s=N_t=N_g$ and $N_i=0$, and so  $f_c=f_p=1$ and $f_i=0$. 

Figure~\ref{fig_mock_ccp} shows the completeness, contamination, and purity 
for the mock groups of different masses. The 2MRS and 6dFGS mock
samples appear to have better membership assignments than the deeper SDSS 
and 2dFGRS mock samples. This happens because, in the two shallower 
samples, larger fractions of groups have a single member galaxy, 
which by definition have perfect completeness and zero contamination. 
For the 2MRS and 6dFGS mocks, $\sim 90\%$ of all groups have 
completeness $\sim 100\%$, being lower for more massive halos. 
For the SDSS and 2dFGRS, about $85\%$ ($95\%$) of the groups have  
completeness $\geq 95\%$ ($\sim 70\%$). On the other hand, about 
$95\%$ and $90\%$ of the groups have zero contamination
for the shallower two and deeper two surveys, respectively.
Overall $80-90\%$ of the groups have purity between $0.95$ and $1.05$,
indicating that there is only a $5\%$ difference in the total number 
of members between the true and selected memberships. 
We also check the global completeness of the identified groups as 
a function of redshift, and the results are shown in 
Figure~\ref{fig_mock_comp}. As expected, it declines
beyond the redshift to which halos of a given mass is complete. 

\subsection{Comparison with other group finders}
\label{ssec_comparisonGF}
%%%%% slim revised
%%%%% HJM revised

As mentioned above, our group finder is built upon the group 
finders of Y05 and L16, but there are differences in details, 
especially in the halo mass proxies. Here we compare the performance of 
our group finder with respect to the earlier group finders by applying 
them to the same mock samples.

%%%%%%%%%%% figure13
\begin{figure*}
\includegraphics[width=0.75\linewidth]{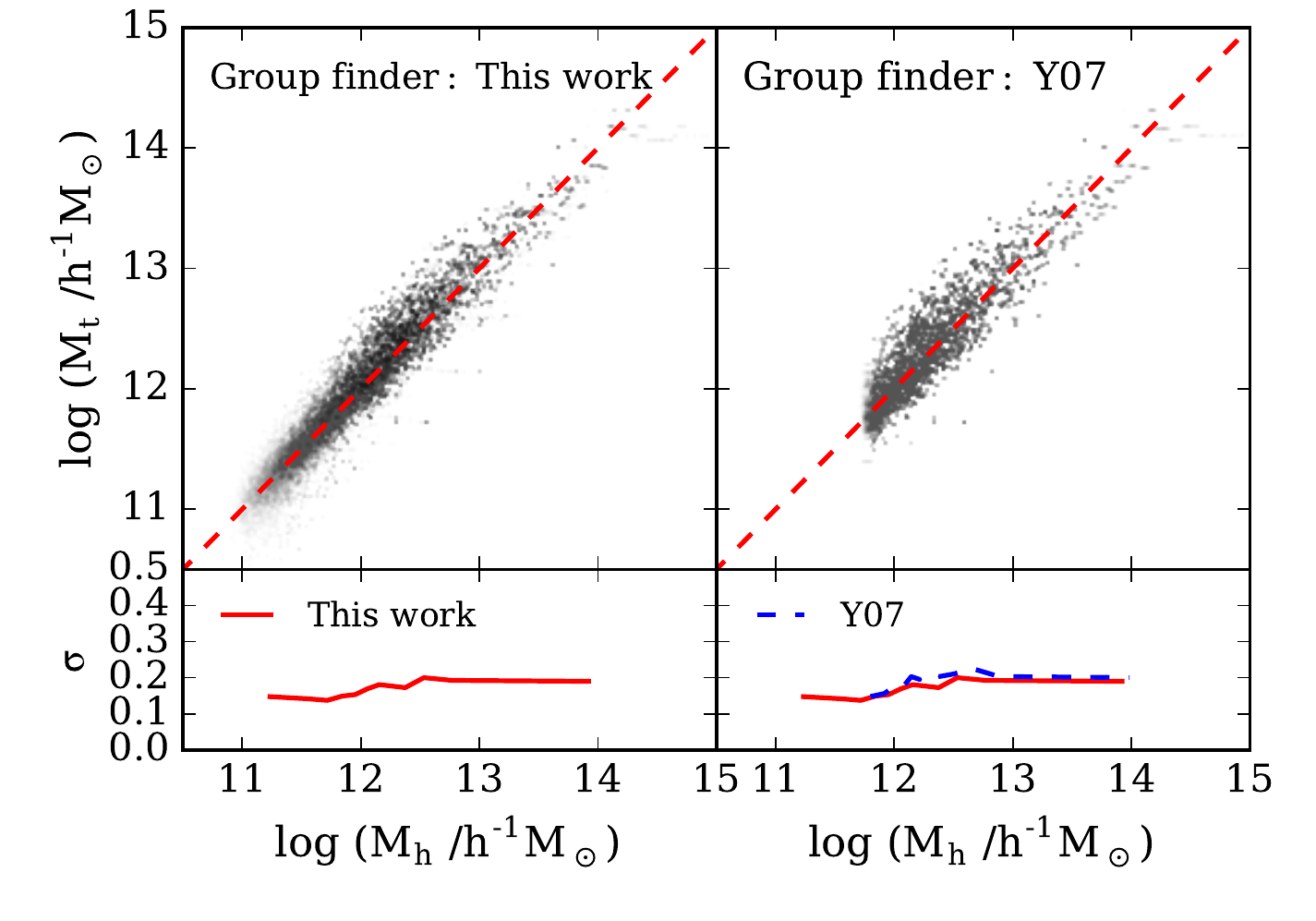}
\caption{Comparison between the true halo mass of the EAGLE simulation and 
the group mass identified by our group finder (left) and by the group finder of 
\citet{yang07} (Y07; right) for the SDSS mock samples restricted to
$z\leq0.09$. The lower panels plot the scatter of the true halo masses
at a given group mass. }
\label{fig_comp_Y07_mass}
\end{figure*}

%%%%%%%%%%% figure14
\begin{figure*}
\includegraphics[width=1.0\linewidth]{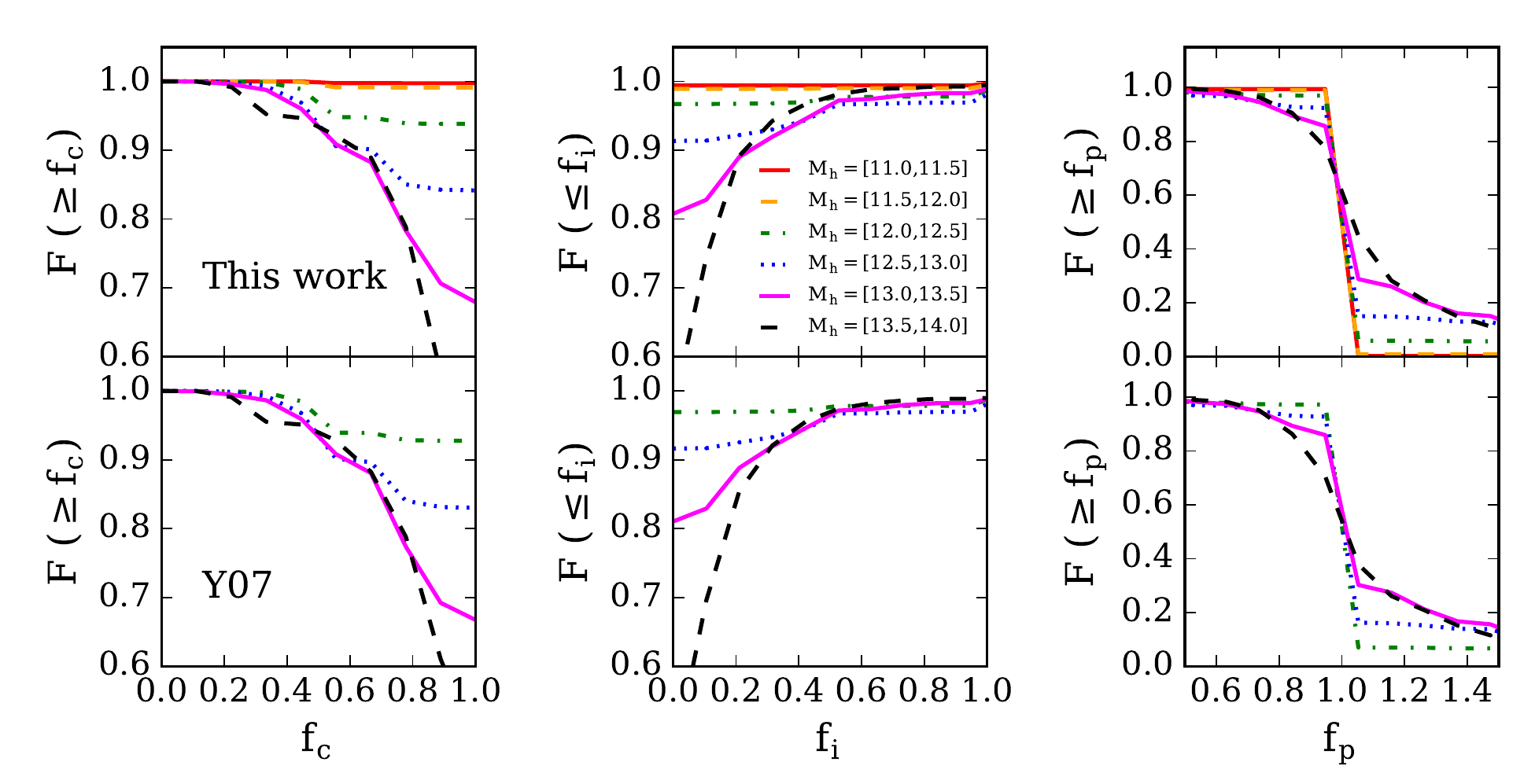}
\caption{Comparison of membership assignment in terms of the completeness (left), 
contamination (middle), and purity (right) between our group finder (upper) and 
the group finder of \citet{yang07} (Y07; lower) for the same mock samples
as Figure~\ref{fig_comp_Y07_mass}. The vertical axis plot the cumulative fraction
of the groups, and the different lines are for haloes of 
different masses as indicated.}
\label{fig_comp_Y07_ccp}
\end{figure*}

%%%%%%%%%%% figure15
\begin{figure*}
\includegraphics[width=0.75\linewidth]{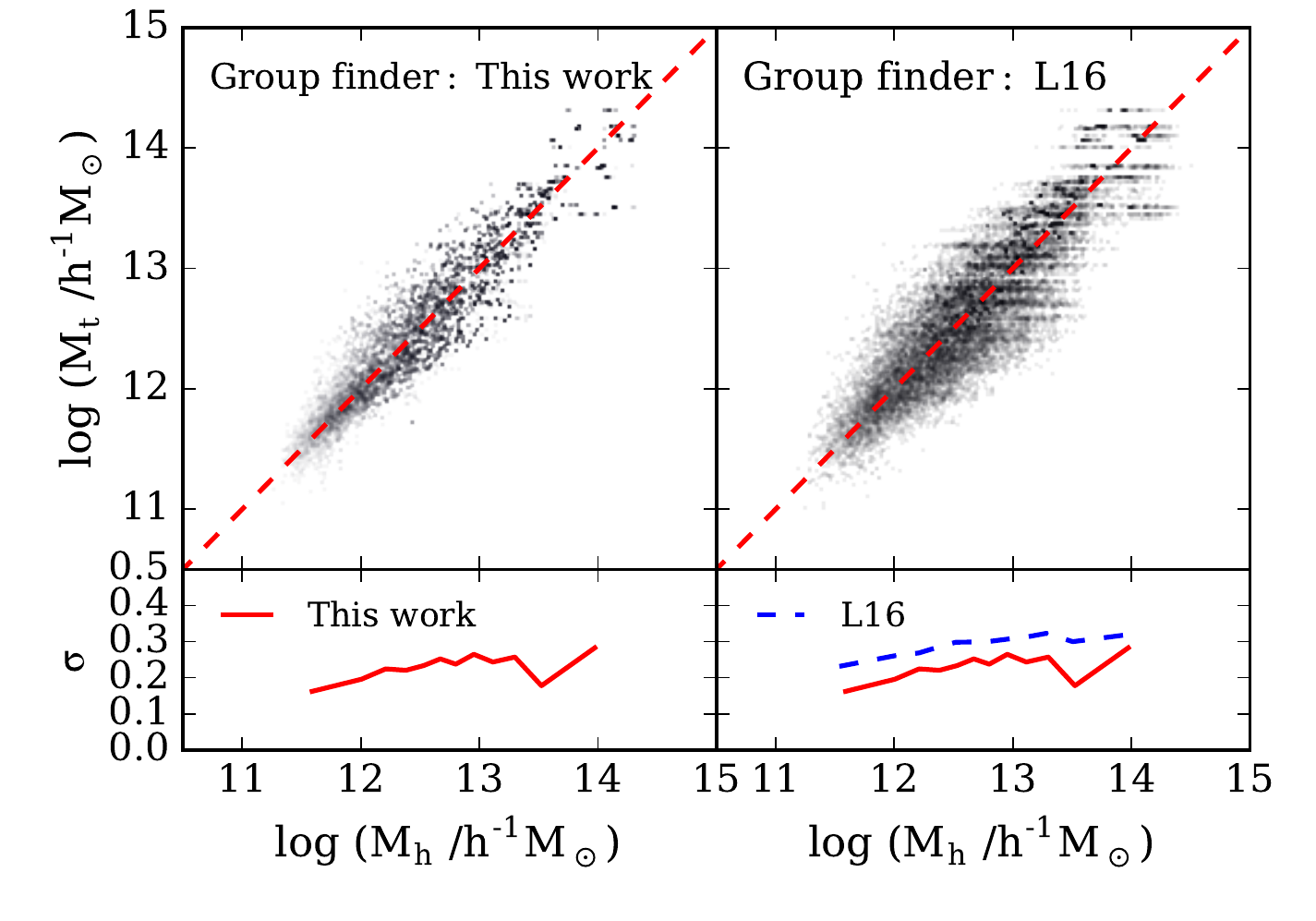}
\caption{Comparison between the true halo mass of the EAGLE simulation and 
the group mass identified by our group finder (left) and by the group finder of 
\citet{lu16} (L16; right) for the 2MRS mock samples. The lower 
panels plot the scatter of the true halo masses at a given group mass. }
\label{fig_comp_L16_mass}
\end{figure*}

\subsubsection{Comparison with Yang et al.}

In \citet{yang07} (Y07), the total group luminosity (group stellar mass) 
of member galaxies brighter than $M_r=-19.5+5\log (h)$ in the $r$-band was 
used as the proxy of the halo mass. These group luminosity and stellar mass 
will be denoted as $L_{19.5}$ and $M_{*,19.5}$, respectively.  
For galaxies at redshifts where the survey limit corresponds to an 
absolute magnitude brighter than the limit, Y07 used the observed 
luminosity function to account for the contribution to $L_{19.5}$
from the missing galaxies due to the magnitude limit. However, 
as found in Y07, this correction introduces uncertainties in the 
group masses. This is not surprising given that most groups identified have 
only a few member galaxies even for the SDSS, and an extrapolation 
according to an average luminosity function is not expected to 
give an accurate estimate of $L_{19.5}$ ($M_{*,19.5}$) for individual 
groups. Y07 found that the uncertainty introduced by this is larger 
than that introduced by the group finder itself, and is comparable to 
the intrinsic scatter in the true halo mass at a given $L_{19.5}$ 
(or $M_{*,19.5}$). As a more demanding test of our group finder 
against that of Y07, we restrict the mock sample to $z\leq\sim0.09$, 
the redshift limit to which the selection is complete to $M_r=-19.5+5\log (h)$ 
so that no extrapolation is needed in the group mass proxy used 
in Y07. The mock sample here is that constructed for the SDSS 
from the EAGLE simulation, as described in \S\ref{ssec_mock}.

Figure~\ref{fig_comp_Y07_mass} shows the group masses 
obtained from our group finder and the Y07 group finder 
with respect to the  true halo masses of the simulation. 
When applying the Y07 group finder, the ranking of groups in 
$M_{*,19.5}$ is used to assign group masses, while our group 
finder uses the halo mass proxy (stellar mass based)
as described in \S\ref{ssec_massproxy}. As one can see, our
group finder matches the true halo masses with an accuracy 
slightly higher than that of Y07, with scatter typically  
of $0.15-0.2\,{\rm dex}$. This indicates that Proxy-M and the 
GAP correction work as well as using $M_{*,19.5}$ to assign 
halo masses to groups. However, had we included groups 
at $z>0.09$, where extrapolation is needed in Y07's group mass
proxy, the scatter given by Y07 would become $0.25-0.3\,{\rm dex}$ 
while that given by our group finder remains at the level 
of $0.2\,{\rm dex}$. 
In addition, our group finder performs equally well 
even for halos with masses as low as $10^{11} h^{-1}{\rm M}_\odot$,   
about an order of magnitude lower than that reached by Y07.
Many of these low mass halos contain only galaxies with 
$M_r$ fainter than $-19.5+5\log h$, which are not assigned 
halo masses in the original Y07 method. 
However, while the group finder of Y07 itself 
does not include these low-mass groups in the SDSS group catalog,  
halo mass assignment can be extended to lower masses by using a 
relation between halo mass and central galaxy, as given in, e.g.,  
\citet[][]{yang12}. The number of groups identified by our group finder 
and the Y07 group finder are $180,835$ and $184,833$, of which $35,376$ and 
$32,343$ have more than one member, respectively. These are very 
close to the true number of halos of $177,013$, of which 
$35,439$ have more than one member. 

Figure~\ref{fig_comp_Y07_ccp} shows the comparison of the two group 
finders in group completeness, contamination, and purity. 
For both of the group finders, the completeness decreases 
and the contamination increases with increasing halo masses, 
as we have seen in \S\ref{ssec_purity}. The two group finders perform 
almost equally well in membership assignments.  

\subsubsection{Comparison with Lu et al.}

L16 developed and calibrated their group finder with their 2MRS mock samples
constructed from an empirical conditional luminosity function model 
(see L16 for details). Here we apply our group finder and that of L16 to 
our own 2MRS mock sample, and make comparisons in their performances. 
For both of the group finders, we adopt the functional forms given by 
equation (9) and equation (10), and the corresponding best parameters 
for the GAP correction for groups of more than one member. 
Otherwise we follow the methodology of L16 as closely 
as possible to reproduce their group finder. The major
difference between the two group finders is in the prescription for 
isolated galaxies. While our group finder uses Proxy-L to assign 
halo masses, we follow the prescription of L16 for their group
finder. There are a total of $29,464$ true halos, of which $5,158$ 
have more than one member, and the number of groups identified by our 
group finder (by L16) are $30,118$ ($29,968$), of which $4,980$ ($4,522$) have more
than one member. Furthermore, our group finder identifies 
$879$ groups with three members and $364$ with four members. 
The corresponding numbers by L16 are $888$ and $362$. 

The group masses obtained by the two group finders are compared to
the true halo masses in Figure~\ref{fig_comp_L16_mass}. One can see that 
our group finder reduces the overall scatter by $\sim0.1\,{\rm dex}$ 
relative to that given by L16. As the two group finders work in a similar way for groups 
of more than one member, the improvement in our group finder is mainly 
due to a better mass proxy for groups containing only one member. 
Note also 
that the mass proxy used by L16 is calibrated with a mock catalog constructed 
from the observed conditional luminosity functions in the $r$ band and scaled 
to the $K$ band using abundance matching, while our mass proxies are 
calibrated with the EAGLE simulation. Part of the difference may also 
be due to the different calibrations. The scatter we obtain here
for the L16 group finder is very similar to that obtained in the 
original L16 paper from a completely different mock sample, suggesting that 
the test results are not particularly sensitive to the mock samples 
adopted for the test. This is also demonstrated in Appendix A, 
where it is shown that our group finder performs equally well for an 
independent mock sample constructed from an empirical model of galaxy formation.

%%%%%%% SECTION 5
\section[gcatalog]{THE GROUP CATALOGS}
\label{sec_gcatalog}

%%%%%%%%%%% figure16
\begin{figure*}
\includegraphics[width=0.9\linewidth]{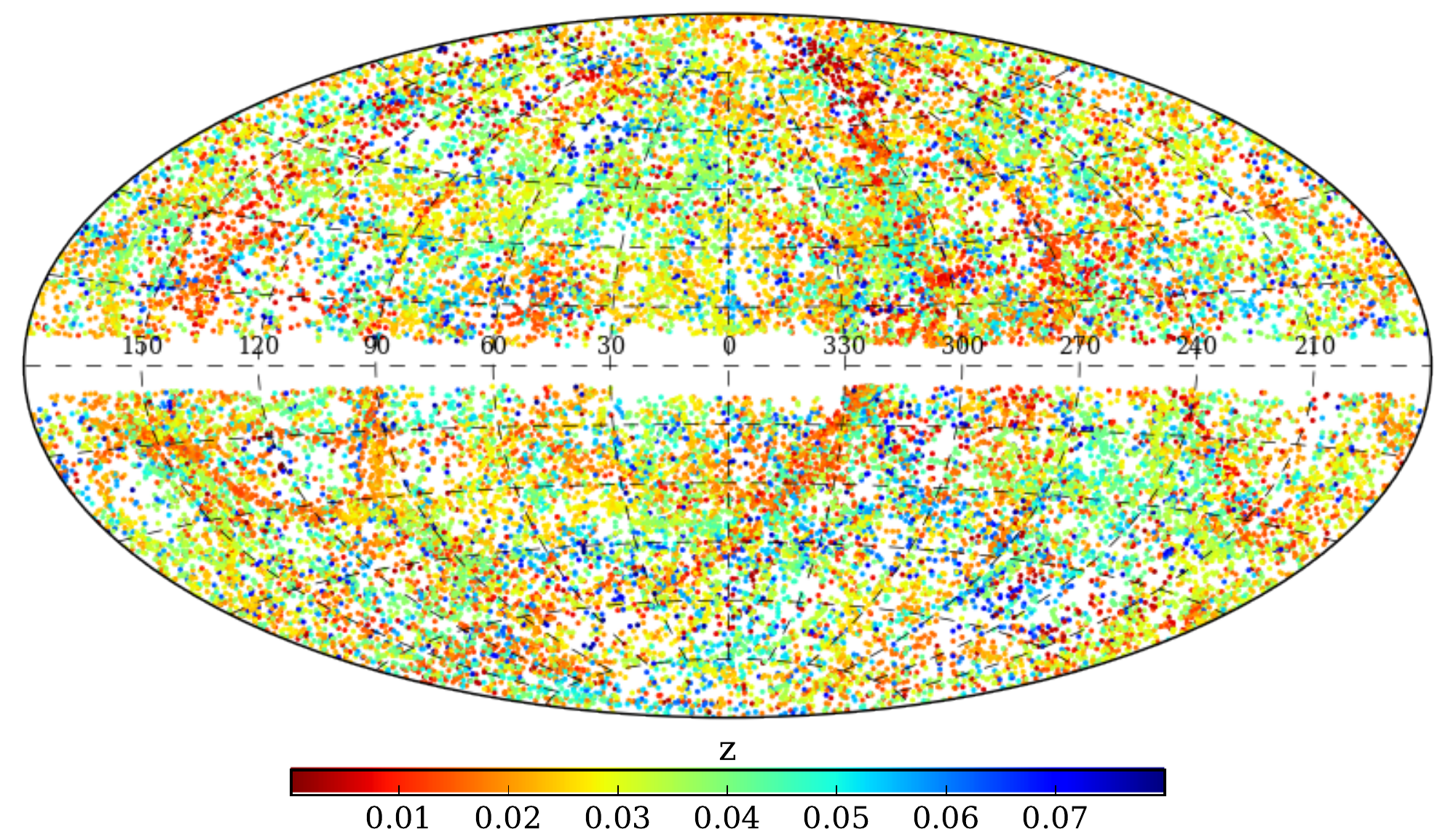}
\caption{The group distributions in Galactic coordinates (Aitoff projection) 
of the 2MRS group catalog.}
\label{fig_2MRS_dist}
\end{figure*}

%%%% Table ?? %%%%
\begin{table*}
 \renewcommand{\arraystretch}{1.5} 
 \centering
 \begin{minipage}{173mm}
  \caption{A summary of group catalogs.}
  \begin{tabular}{cccccccc}
\hline
Catalog & Total galaxies & Total groups\textsuperscript{\footnote{This includes groups 
without reliable halo mass assigned because halo mass is not complete at given 
redshift.}} & Total groups & $N\textsuperscript{\footnote{The number of member galaxies 
in a group. This excludes groups without reliable halo mass assignment.}}=1$ & $N\geq2$ & 
$M_h\geq10^{14}{\rm M_\odot}/h$ & $M_h\geq10^{13}{\rm M_\odot}/h$ \\
 &  &  & with reliable mass & & & &  \\
\hline
\hline
2MRS(L)       & $43,249$ & $30,937$ & $18,650$ & $13,311$ & $5,339$ & $982$ & $6,836$ \\
2MRS(M)       & $43,249$ & $31,752$ & $19,224$ & $13,913$ & $5,311$ & $1,016$ & $7,156$ \\
2MRS$+$(L)    & $44,310$ & $31,804$ & $18,731$ & $13,275$ & $5,456$ & $984$ & $6,877$ \\
2MRS$+$(M)    & $44,310$ & $32,693$ & $19,307$ & $13,923$ & $5,384$ & $1,014$ & $7,211$ \\

6dFGS(L)     & $62,987$ & $46,676$ & $17,907$ & $11,126$ & $6,781$ & $1,004$ & $6,919$ \\
6dFGS(M)     & $62,987$ & $47,176$ & $18,555$ & $11,789$ & $6,766$ & $1,045$ & $7,291$ \\
6dFGS$+$(L)  & $73,386$ & $59,515$ & $21,481$ & $14,168$ & $7,313$ & $1,154$ & $8,030$ \\
6dFGS$+$(M)  & $73,386$ & $59,512$ & $22,223$ & $15,278$ & $6,945$ & $1,191$ & $8,459$ \\

SDSS(L)      & $586,025$ & $446,495$ & $165,538$ & $112,444$ & $53,094$ & $3,757$ & $39,565$ \\
SDSS(M)      & $586,025$ & $421,715$ & $167,638$ & $105,979$ & $61,659$ & $3,780$ & $43,880$ \\
SDSS$+$(L)   & $600,458$ & $453,927$ & $164,694$ & $107,066$ & $57,528$ & $3,712$ & $39,464$ \\
SDSS$+$(M)   & $600,458$ & $426,932$ & $166,999$ & $101,518$ & $65,481$ & $3,760$ & $43,649$ \\

2dFGRS(L)    & $180,967$ & $144,965$ & $77,423$ & $62,101$ & $15,322$ & $606$ & $8,526$ \\
2dFGRS(M)    & $180,967$ & $145,756$ & $77,365$ & $61,309$ & $16,056$ & $632$ & $9,116$ \\
2dFGRS$+$(L) & $189,101$ & $147,757$ & $77,861$ & $59,606$ & $18,255$ & $634$ & $8,553$ \\
2dFGRS$+$(M) & $189,101$ & $148,290$ & $77,757$ & $58,909$ & $18,848$ & $638$ & $9,099$\\
\hline
\vspace{-2mm}
\end{tabular}
\textbf{Notes.}
\vspace{-5mm}
\label{tab_catalog}
\end{minipage}
\end{table*}
%%%%%%%%%%%%%%%%%%%%%%%%%%%%%%%%%%%%%%%%%%%%%%%%%%%%%%%%%%%%%%%%%%%%%%%%%%%%%%

%%%%%%%%%%% figure17
\begin{figure*}
\vspace{-0.43cm}
\includegraphics[width=0.75\linewidth]{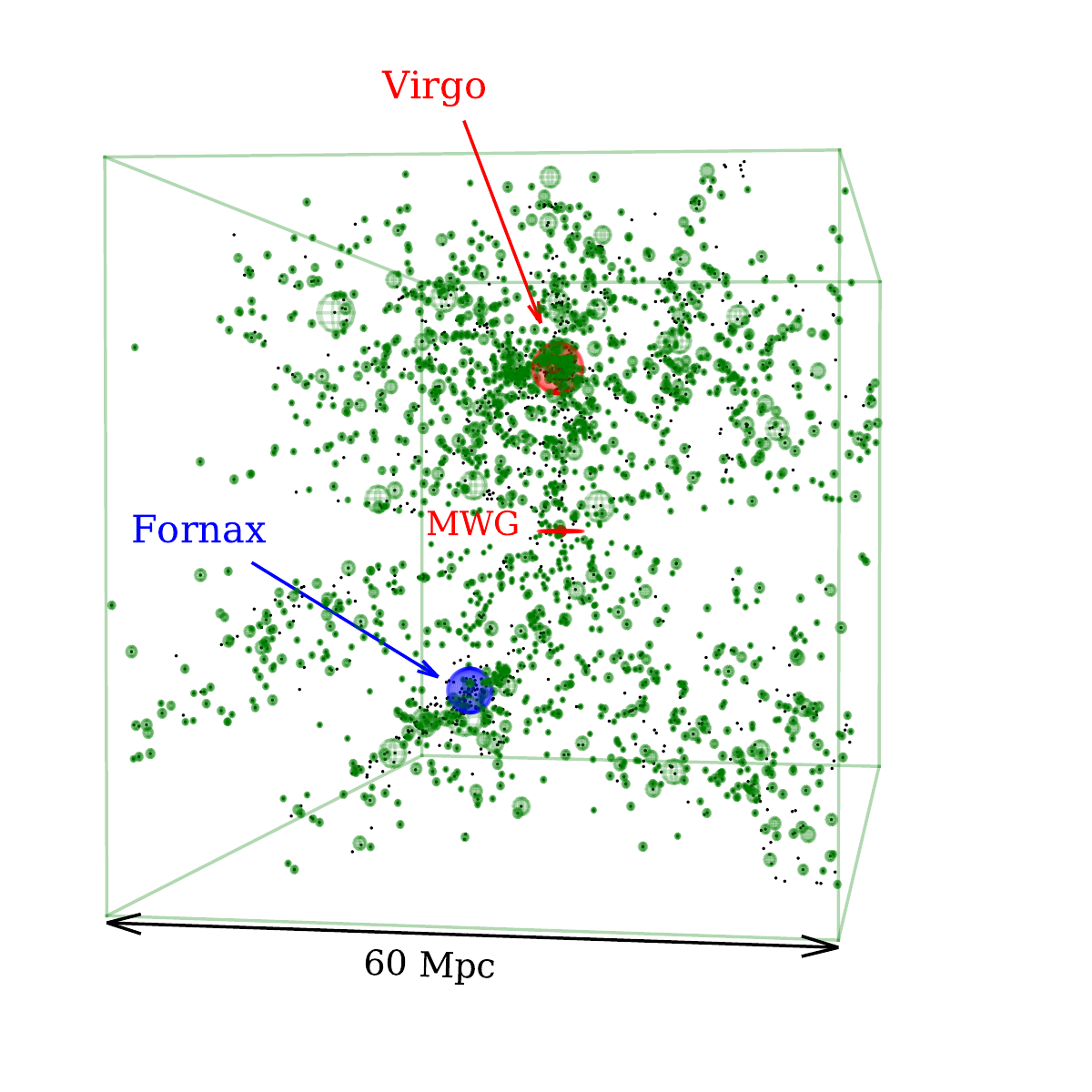}
\vspace{-1cm}
\hspace{-2cm}
\caption{Three dimensional distribution of the 2MRS galaxies (black dots) 
and groups identified by the group finder (wire-framed green spheres with 
radii of $r_{180}$) in the local Universe with the Milky Way at the center.}
\label{fig_3D}
\end{figure*}

%%%%%%%%%%% figure18
\begin{figure*}
\includegraphics[width=0.95\linewidth]{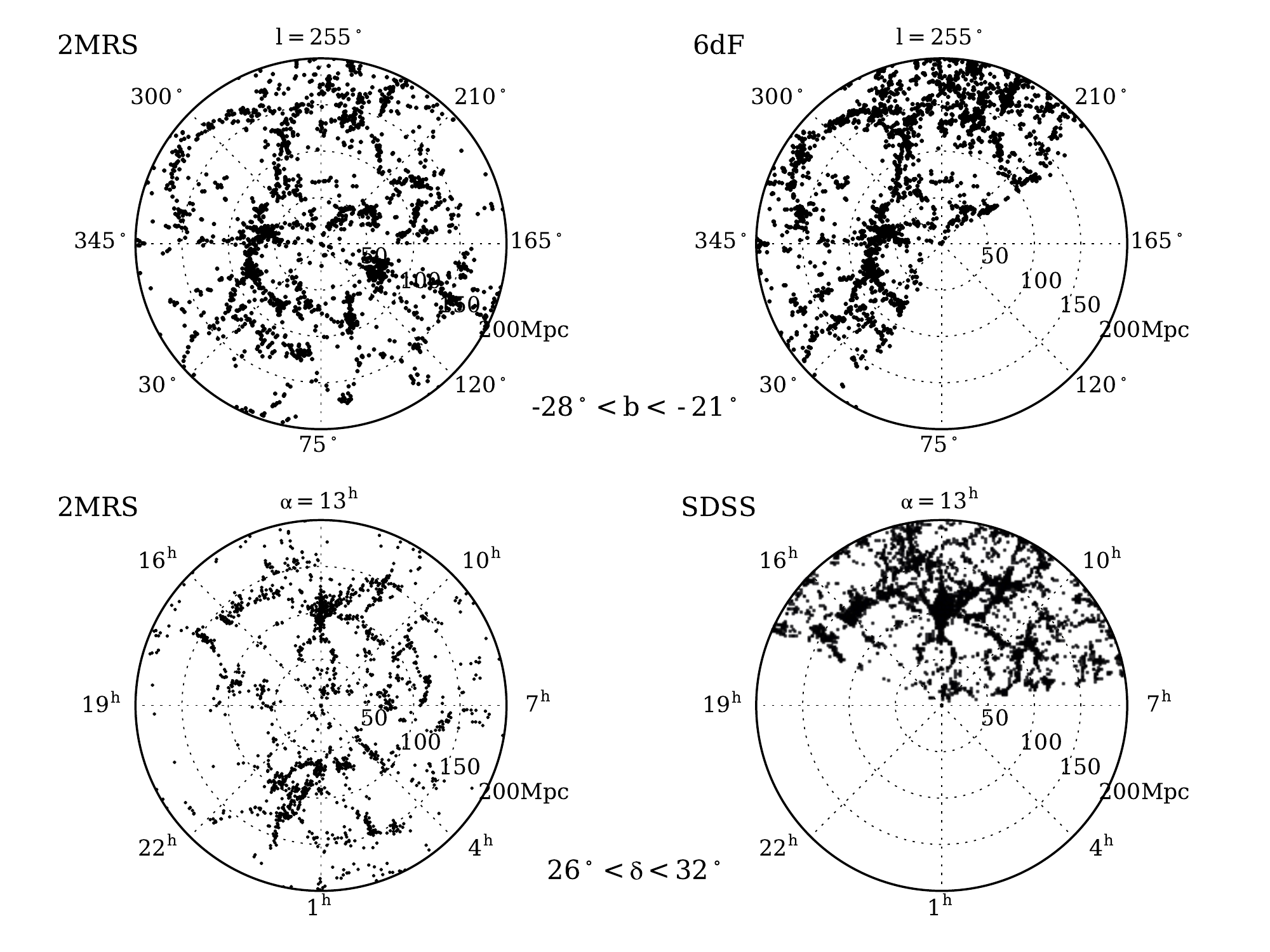}
\caption{Comparison of the distribution of the groups constructed by 
the group finder in slices between different surveys. The upper and the lower
panels show different slices as indicated.}
\label{fig_slice}
\end{figure*}

%%%%%%%%%%% figure19
\begin{figure*}
\includegraphics[width=0.75\linewidth]{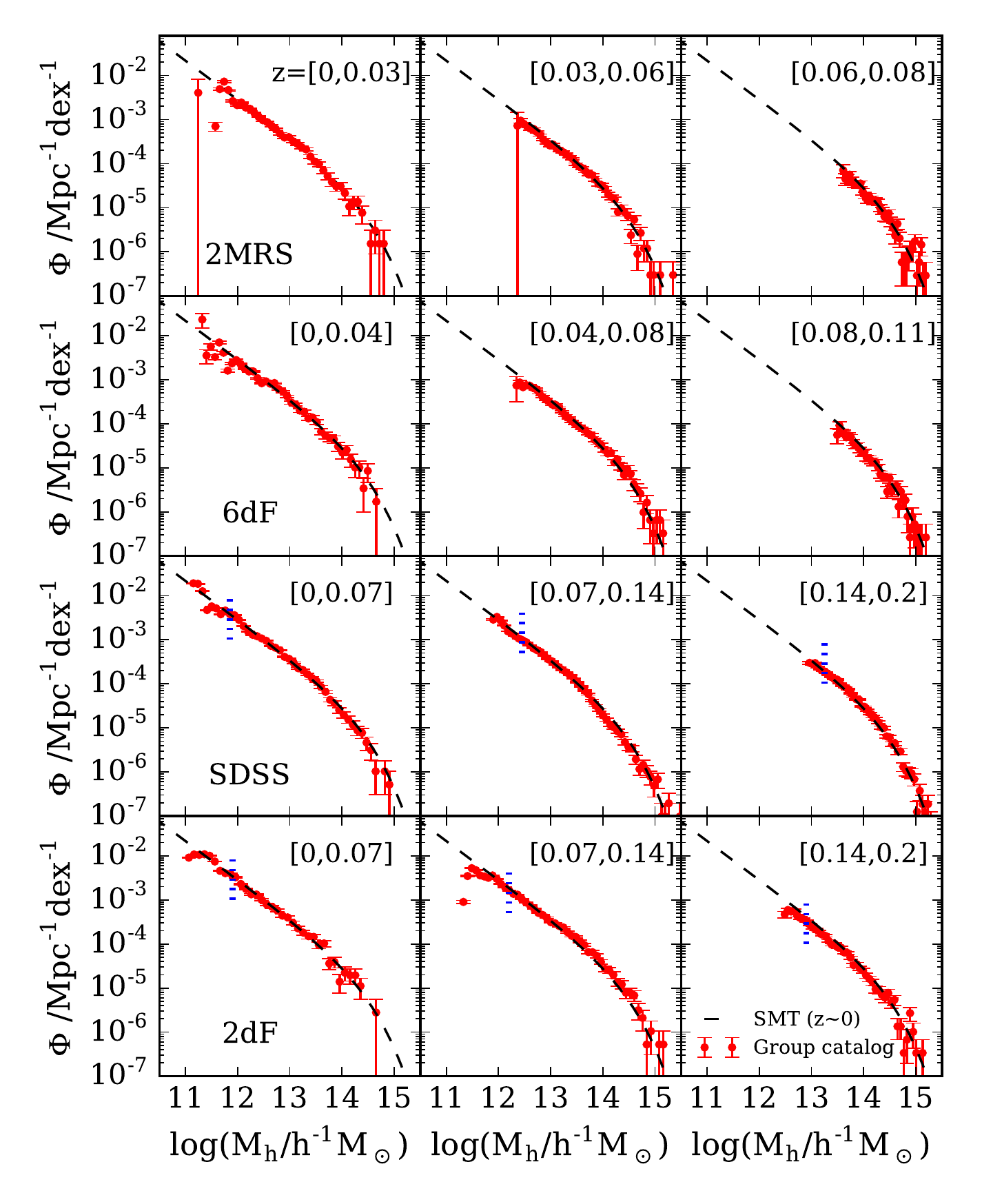}
\caption{Halo mass functions of galaxy systems constructed based on the
four surveys. The dashed lines are the theoretical mass function by 
\citet{sheth01}, which we used for abundance matching for the group finder. 
The dotted short ticks in the lower two panels indicate the lower limits of 
the halo masses of the group catalogs by \citet{yang07} and \citet{yang05}, 
respectively.}
\label{fig_hmf_comp}
\end{figure*}

In this section, we present the group catalogs we construct by applying the group 
finder, as described in \S\ref{sec_gfinder}, to the observational 
samples described in \S\ref{sec_data}. As mentioned earlier,
we provide four catalogs for each observation sample:  
\begin{enumerate}
    \item a catalog constructed with galaxies that have spectroscopic redshifts, 
       using Proxy-L to estimate halo masses;
    \item a catalog constructed with galaxies that have  spectroscopic redshifts, 
       using Proxy-M to estimate halo masses;
    \item a catalog constructed with all galaxies, using Proxy-L to estimate halo masses;
    \item and a catalog constructed with all galaxies, using Proxy-M to estimate halo masses.
\end{enumerate}
For convenience we will use the name of the galaxy sample together with 
the halo mass proxy 
adopted to refer to a group catalog. For example, the catalogs constructed 
from the SDSS survey are referred to as SDSS(L), SDSS(M), SDSS$+$(L), 
and SDSS$+$(M), respectively. For brevity, the following presentations 
are mainly based on catalogs (i) and (ii), unless stated otherwise.

\subsection{The catalogs and their basic properties}
%%% slim revised
%%% HJM revised. 

For SDSS and 2dFGRS, Proxy-L uses the $r$- and $R$-band luminosities, respectively, 
and Proxiy-M uses the stellar masses of galaxies as described in 
\S\ref{ssec_SDSScat} and \S\ref{ssec_2dFGRScat}.
For 2MRS and 6dFGS, Proxy-L is based on the $K_s$-band luminosities of galaxies.
The Proxy-M for these two samples are based on the stellar masses obtained from 
the mean relation between the $K_s$-band luminosity and stellar mass from the EAGLE 
simulation. We use the same calibrations of the halo mass proxies as described in 
\S\ref{ssec_massproxy}. Our tests show that it is not necessary to re-calibrate 
the mass proxies for individual samples, as the outcomes with and
without such a re-calibration converge in the end. This is expected, 
because our group finder uses the mass proxies only to rank group
masses, and the halo masses are re-adjusted at the end of each iteration 
using abundance matching. In the tables and figures 
shown in this section, we exclude groups that are not assigned halo masses 
by abundance matching because of sample incompleteness at the given halo mass 
and redshift (see \S\ref{ssec_mass}). In the catalogs, however, we include 
these groups (with a flag), and  assign them masses according to the mean relation between 
the halo mass and the mass proxy obtained from the last iteration of the group 
finder. 

The distribution in the sky of the 2MRS groups selected by our group 
finder is shown in Figure~\ref{fig_2MRS_dist}, and 
Figure~\ref{fig_3D} shows the three dimensional distributions of galaxies and 
groups in the local Universe from the 2MRS catalogs. Also, the distributions of 
groups from different surveys in the same slices are compared in 
Figure~\ref{fig_slice}. 

Figure~\ref{fig_hmf_comp} compares the mass function of the halos from the 
group catalogs with the theoretical model of \citet{sheth01}, which was
used for abundance matching for the group finder. The good agreement between 
the observational data and the theoretical model is largely by design. 
However, the plots do show the halo-mass and redshift ranges covered by different 
samples, as well as the statistical uncertainties in the number densities of groups.  

\begin{figure*}
\includegraphics[width=0.75\linewidth]{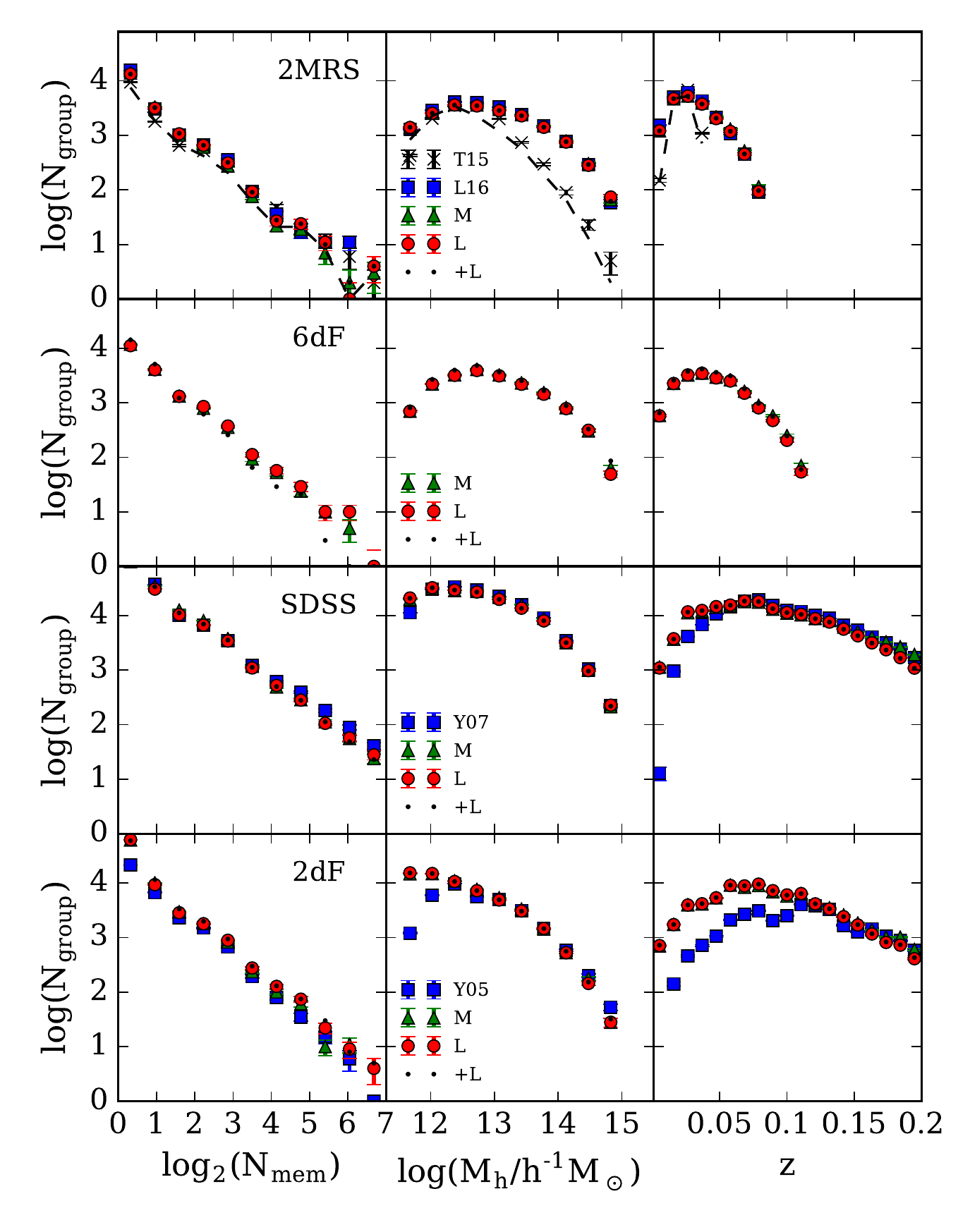}
\caption{The number of groups as a function of the number of members,
halo mass, and redshift for L (circle) and M (triangle) catalogs for each survey, 
compared with \citet{tully15} (T15; cross) and \citet{lu16} (L16; square) 
for the 2MRS, with \citet{yang07} (Y07; square) for the SDSS, and with 
\citet{yang05}(Y05; square) for the 2dFGRS. The results from the $+$(L) catalogs 
(dots) are also shown for comparison. The T15 results (crosses)
should be compared with the dashed lines, which are obtained by using  
only groups with recession velocities between $3,000$ and 
$10,000{\rm km\ s^{-1}}$, within which the T15 sample is complete. 
The comparison is only made for groups that halo mass is complete at a given redshift. 
The error bars shown represent Poisson errors.}
\label{fig_real_comp}
\end{figure*}

Table~\ref{tab_catalog} lists the total number of groups, as well as the number 
of groups of given richness and halo masses selected from different samples. 
Figure~\ref{fig_real_comp} shows in more detail the distributions 
of groups with respect to richness (number of member galaxies), 
halo mass, and redshift. Note, again, that these distributions are obtained 
from groups for which halo masses are complete at a given redshift, 
as shown in Figure~\ref{fig_mock_Mcomp_z}. It is seen that the 
results from (L) and (M) catalogs are consistent with each other 
within the Poisson uncertainties. For comparison, we also show 
the results for the $+$(L) catalogs [the results for the 
$+$(M) catalogs are very similar] as the small dots. As one can see,
results from the extended ($+$) catalogs are consistent with the non-extended
catalogs, except for the 6dFGS which has the poorest completeness in 
spectroscopic redshifts. Also, some massive clusters in 
the catalogs have only one galaxy particularly for 2MRS 
and 6dFGS, because of their shallow depths that make satellites not
observable. 

\begin{figure*}
\includegraphics[width=0.75\linewidth]{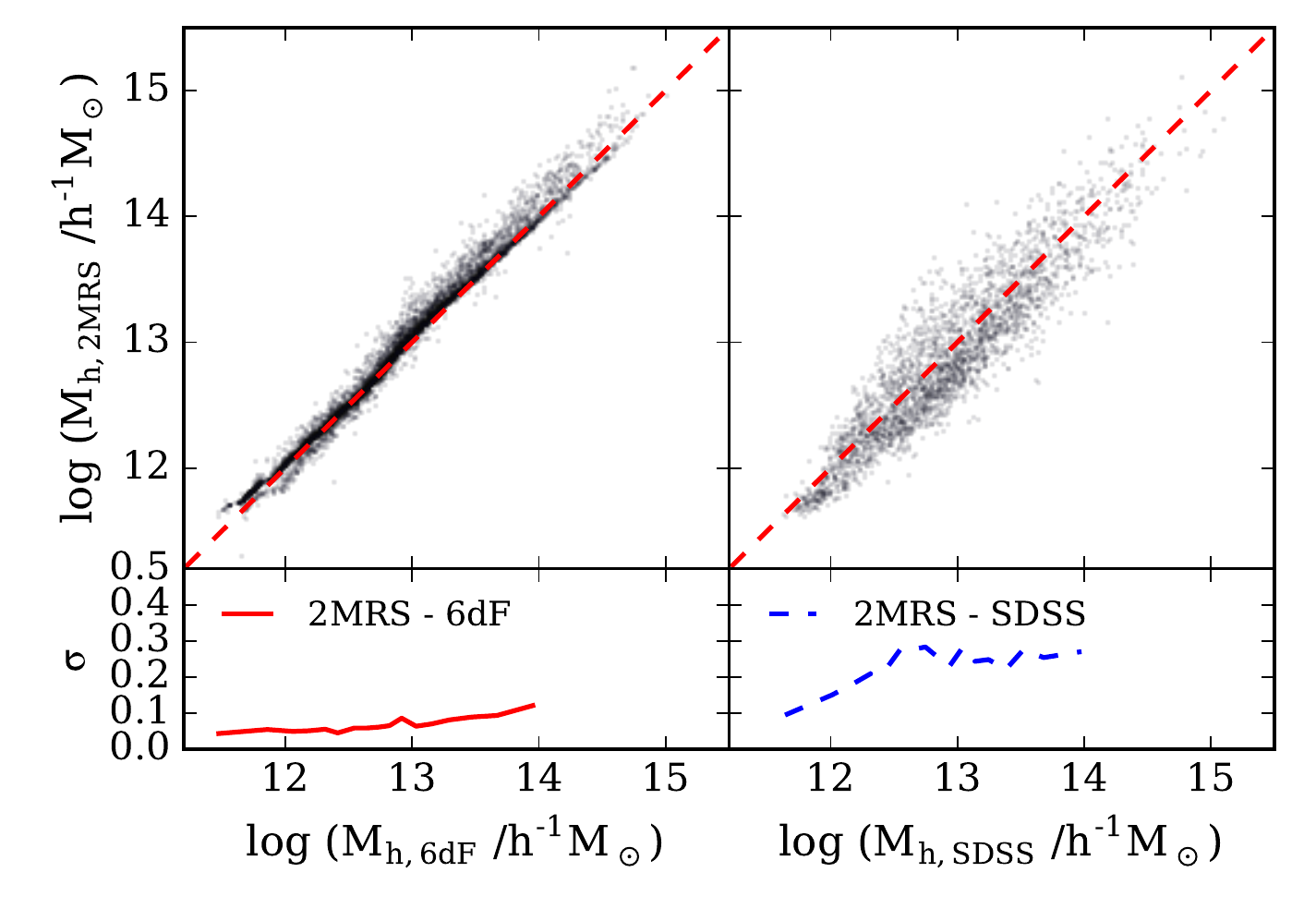}
\caption{Comparison of halo masses (based on Proxy-M) for individual groups 
cross-identified between different group catalogs. We used the tolerances of 
$\leq10\arcsec$ and $|\Delta z|\leq 10^{-3}$ for the cross-identification. 
The lower panels plot the scatter of the 2MRS halo masses at a given mass from 
the other catalogs.}
\label{fig_real_comp_mass}
\end{figure*}

Figure~\ref{fig_real_comp_mass} compares halo masses (based on Proxy-M) 
for individual groups cross-identified between the group catalogs. 
While we do not present the comparison for the 2dFGRS because the number 
of such groups is small, we did check that the mean relation and scatter for 
the 2dFGRS are similar to those for the SDSS. 
We used the tolerances of angular separation less than $10\arcsec$ and 
$|\Delta z|\leq 10^{-3}$ for the cross-identifications. One can see that there 
is a very tight correlation in halo masses between the 2MRS and the 6dFGS
group catalogs, while a larger dispersion of $0.2-0.3\,{\rm dex}$ is found 
between the 2MRS and SDSS, mainly because of the differences in the stellar mass 
estimates.

%%%%%%%%%%% figure20

%%%%%%%%%%% figure21

\subsection{Comparison with other catalogs}

Here we compare our catalogs with a number of other catalogs in the literature, 
including the 2MRS catalogs of L16 and \citet{tully15} (T15), the SDSS catalog of Y07, 
and the 2dFGRS catalog of Y05. 

\subsubsection{Comparison of the 2MRS group catalog with L16}
%%%%HJM revised

As mentioned earlier, L16 built a group catalog based on the same sample of 
the 2MRS galaxies as ours using a similar methodology. 
Figure~\ref{fig_real_comp} shows the comparison 
between the two catalogs in the number of groups as functions of 
richness, mass, and redshift, and it is clear that the two catalogs are in 
good agreements. We also checked the mass of individual haloes 
for groups that are cross-identified between the two catalogs, and found 
that our mass assignments are in general agreement with those of L16 with 
a typical dispersion of $\sim0.25\,{\rm dex}$ between them. 

As described earlier, our group finder is different from that of L16
in two ways. The first is that we re-calibrated the gap-based 
mass model of L16, so that the mass assigned to a group may be different 
from that of L16 even if it has the same membership of galaxies.  
Second, L16 used a `Gap limit' prescription to assign masses 
for groups containing only one member (see \S\ref{ssec_proxyforpoor}), 
while our group finder does not. We believe that these 
are the sources of the dispersions and discrepancies found
in the comparison of halo mass between the two catalogs. 

\subsubsection{Comparison of the 2MRS group catalog with T15}
%%%%HJM revised

\citet{tully15} (T15) constructed a 2MRS group catalog using an empirical 
relation between halo mass (and the corresponding size and 
velocity dispersion) and a characteristic group luminosity to assign 
galaxies into groups. Figure~\ref{fig_real_comp} shows the comparison
of our catalog with theirs in the number of groups as functions 
of richness, mass, and redshift. As T15 stated that their 
group catalog is less reliable outside the recession velocity range 
between $3,000$ and $10,000{\rm km\ s^{-1}}$, we 
make comparisons only for the groups 
within the velocity range. One can see that T15 contains more massive
clusters than our catalogs, 
while the richness and redshift distributions are in  better agreements. 
T15 compared the mass function of 
their groups with a theoretical halo mass function and found 
that, although the shape of their group mass function 
is similar to that of the theoretical function, the normalization
is about a factor of $4.6$ higher. As mentioned above, T15 used an 
empirical model for their group masses, which is different from the 
mass proxies used in our group finder. Furthermore, their definition
of halo masses is also different from ours. All these produce the 
differences seen between the two catalogs.

\subsubsection{Comparison of the SDSS group catalog with Yang et al.}

Y07 built a group catalog of the SDSS DR7 galaxies. As we described earlier, 
their group finder is similar to ours in that it uses halo mass and velocity
dispersion of groups identified to update galaxy memberships at each iteration 
until its iteration reaches convergence in the membership assignments. 
The main difference is that it uses a summed stellar mass (or luminosity) of 
member galaxies brighter than $M_{r}=-19.5+5\log h$ as a halo mass proxy. 
However, as the group catalogs are dominated by 
groups containing one galaxy (see Table~\ref{tab_catalog}), it results in no 
significant net difference except that our group catalog extends to lower mass. 
Figure~\ref{fig_real_comp} compares the SDSS group catalog of Y07 with that  
given by our group finder. The Y07 catalog contains smaller number of low-mass
systems at relatively low redshifts, mainly because of the magnitude limit of 
$M_{r}=-19.5+5\log h$ adopted by Y07 in their halo mass proxy, which is brighter
than the observational flux limit at $z<\sim0.09$.

\subsubsection{Comparison of the 2dFGRS group catalog with Yang et al.}

Y05 also constructed a group catalog of the 2dFGRS galaxies. Again, their group 
finder is similar to ours but differs in that Y05 uses a summed stellar mass of 
member galaxies brighter than 
$M_{b_J}=-18+5\log h$ as a halo mass proxy instead of stellar mass (luminosity) 
of central galaxy and the $n$-th most massive (brightest) galaxy, 
which our group finder uses. The sample selection for the 2dFGRS is almost identical 
to our sample selection.
Figure~\ref{fig_real_comp} shows comparisons between the two catalogs in 
the number of groups of given richness, mass, and redshift. The lower number of 
groups in Y05 in low-mass end and low-redshift is again because of the limit of 
$M_{b_J}=-18+5\log h$ used by Y07 for the halo mass proxy, which is brighter
than the flux limit in the observation at $z<\sim0.12$. 
Otherwise, the agreement between the two catalogs is reasonably good.

\subsection{Contents of the catalogs}
%%%%%HJM revised
%%% slim revised

The galaxy and group catalogs constructed are available at 
{\tt http://gax.sjtu.edu.cn/data/Group.html}.  
The group catalogs list the properties
of groups, while the galaxy samples present not only the properties 
of galaxies but also their links to groups. Object indexes are also 
provided for galaxies so that one can identify them from the original 
galaxy catalogs. As mentioned above, there are four group catalogs 
for each galaxy sample, and so in total we provide 16 group catalogs, 
as summarized in Table~\ref{tab_catalog}. Tables~\ref{tab_2MRS_grp} 
and \ref{tab_2MRS_gal} show the structures of the catalogs we provide, 
using the 2MRS as an example. In what follows we explain the different 
columns in more detail. 

%%%% Table ?? %%%%
\begin{table*}
 \renewcommand{\arraystretch}{1.5} 
 \centering
 \begin{minipage}{153mm}
  \caption{The 2MRS group catalog.\textsuperscript{a}}
  \begin{tabular}{cccccccccc}
\hline
$(1)$ & $(2)$ & $(3)$ & $(4)$ & $(5)$ & $(6)$ & $(7)$ & $(8)$ & $(9)$ & $(10)$ \\
group ID & cen ID & ra & dec & $z$ & $\log(M_h/h^{-1}{\rm M_\odot})$ & 
$N_{\rm mem}$ & $f_{\rm edge}$ & i-o & known as \\
 & & (deg) & (deg) & & & & & &  \\
\hline
\hline

$1$ & $15$ & $187.74899$ & $12.20402$ & $0.00362$ & $14.290$ & $109$ & $1.00$ & $1$ & Virgo \\

$2$ & $486$ & $243.52157$ & $-60.79748$ & $0.01663$ & $14.366$ & $106$ & $1.00$ & $1$ & Norma \\

$3$ & $530$ & $49.47894$ & $41.53527$ & $0.01748$ & $14.297$ & $92$ & $1.00$ & $1$ & Perseus \\

$4$ & $672$ & $194.81308$ & $27.97206$ & $0.02476$ & $14.639$ & $88$ & $1.00$ & $1$ & Coma \\

$5$ & $386$ & $192.33072$ & $-41.18290$ & $0.01437$ & $14.342$ & $62$ & $1.00$ & $1$ & Centaurus\\

$6$ & $1094$ & $258.01088$ & $-23.30095$ & $0.03030$ & $14.758$ & $52$ & $1.00$ & $1$ & Ophiuchus\\

$7$ & $18$ & $53.13952$ & $-35.56552$ & $0.00492$ & $14.097$ & $48$ & $1.00 $& $1$ & Fornax \\

$8$ & $257$ & $210.67554$ & $-33.83448$ & $0.01527$ & $14.534$ & $47$ & $1.00$ & $1$ &  \\

$9$ & $608$ & $17.31215$ & $32.68145$ & $0.01609$ & $14.042$ & $42$ & $1.00$ & $1$ &  \\

$10$ & $432$ & $207.35991$ & $-30.43195$ & $0.01628$ & $14.220$ & $41$ & $1.00$ & $1$ &  \\

\hline
\vspace{-2mm}
\end{tabular}
\textbf{Notes.}

a. The full catalog is available at {\tt http://gax.sjtu.edu.cn/data/Group.html}. 
\vspace{-5mm}
\label{tab_2MRS_grp}
\end{minipage}
\end{table*}
%%%%%%%%%%%%%%%%%%%%%%%%%%%%%%%%%%%%%%%%%%%%%%%%%%%%%%%%%%%%%%%%%%%%%%%%%%%%%%

\subsubsection {The group catalogs}
%%%%%HJM revised

The following items are provided for individual groups.

\begin{DESCRIPTION}

\item[Column (1)]
group ID: an unique ID of a group within a given group catalog;

\item[Column (2)]
cen ID: galaxy ID of the central galaxy of a group
in the corresponding galaxy sample;

\item[Column (3)]
ra (in degrees): right ascension (J2000)
of the luminosity-weighted (for catalogs using Proxy-L) or mass-weighted 
(for catalogs using Proxy-M) group center;

\item[Column (4)]
dec (in degrees): declination (J2000) of the group center;

\item[Column (5)]
$z$: redshift of group center in the CMB rest-frame;

\item[Column (6)]
$\log(M_h/h^{-1}{\rm M_\odot})$: 10-based logarithm of the halo mass of a group  
     in units of $h^{-1}{\rm M}_\odot$;

\item[Column (7)]
$N_{\rm mem}$: number of member galaxies in a group;

\item[Column (8)]
$f_{\rm edge}$: the volume fraction that is not cut out 
from the halo of a group (assumed to be spherical) by the 
survey boundary or mask;

\item[Column (9)]
i-o: A flag that indicates whether a group is inside or outside 
the region of completeness for a given halo mass. 
For a group inside the completeness region (${\rm value}=1$), 
mass is obtained directly from the abundance matching. For a group 
that is outside the completeness region (${\rm value}=0$), mass is estimated 
using the relation between the halo mass and its proxy from the last iteration 
of the group finder. 

\item[Column (10)]
known as: conventional name of a system, identified only for well-known massive clusters. 

\end{DESCRIPTION}

%%%% Table ?? %%%%
\afterpage{
\begin{landscape}
\thispagestyle{empty}
\begin{table}
 \renewcommand{\arraystretch}{1.5} 
 \centering
 \begin{minipage}{233mm}
  \caption{The 2MRS galaxy catalog.\textsuperscript{a}}
  \begin{tabular}{ccccccccccccccc}
\hline
$(1)$ & $(2)$ & $(3)$ & $(4)$ & $(5)$ & $(6)$ & $(7)$ & $(8)$ & $(9)$ & $(10)$ & $(11)$ & $(12)$ & $(13)$ & $(14)$ \\
galaxy ID & survey ID & group ID & ra & dec & $l$ & $b$ & $z_{\rm CMB}$ & $z_{\rm EDD}$ & $z_{\rm comp}$ & $z_{\rm src}$ & 
${\rm dist}_{\rm NN}$ & $\log(L/h^{-2}{\rm L_\odot})$ & $\log (M_*/h^{-2}{\rm M}_\odot)$\\
 & & & (deg) & (deg) & (deg) & (deg) &  & & & & (deg) & \\
\hline
\hline
$1$ & $00424433+4116074$ & $25128$ & $10.68471$ & $41.26875$ & $121.1743$ & $-21.57319$ & $-0.00194$ & $0.00018$ & $1.0$ & $0$ & $0.0$ & $10.644$ & $10.286$ \\
$2$ & $00473313-2517196$ & $1116$ & $11.88806$ & $-25.2888$ & $97.36301$ & $-87.96452$ & $-0.00013$ & $0.00086$ & $1.0$ & $0$ & $0.0$ & $10.771$ & $10.432$ \\
$3$ & $09553318+6903549$ & $345$ & $148.88826$ & $69.06526$ & $142.0919$ & $40.90022$ & $0.00016$ & $0.00085$ & $1.0$ & $0$ & $0.0$ & $10.749$ & $10.406$ \\
$4$ & $13252775-4301073$ & $299$ & $201.36565$ & $-43.01871$ & $309.51639$ & $19.41761$ & $0.00267$ & $0.00083$ & $1.0$ & $0$ & $0.0$ & $10.69$ & $10.339$ \\
$5$ & $13052727-4928044$ & $299$ & $196.36366$ & $-49.4679$ & $305.27151$ & $13.34017$ & $0.00268$ & $0.0$ & $1.0$ & $0$ & $0.0$ & $11.497$ & $11.413$ \\
$6$ & $01335090+3039357$ & $30717$ & $23.4621$ & $30.65994$ & $133.61024$ & $-31.33081$ & $-0.00152$ & $0.00021$ & $1.0$ & $0$ & $0.0$ & $9.439$ & $9.201$ \\
$7$ & $09555243+6940469$ & $345$ & $148.96846$ & $69.6797$ & $141.40953$ & $40.5671$ & $0.00094$ & $0.00082$ & $1.0$ & $0$ & $0.0$ & $10.394$ & $10.02$ \\
$8$ & $03464851+6805459$ & $29389$ & $56.70214$ & $68.09611$ & $138.17259$ & $10.57999$ & $-0.0002$ & $0.00053$ & $1.0$ & $0$ & $0.0$ & $10.12$ & $9.754$ \\
$9$ & $13370091-2951567$ & $5266$ & $204.25383$ & $-29.86576$ & $314.58353$ & $31.97269$ & $0.00263$ & $0.00115$ & $1.0$ & $0$ & $0.0$ & $10.691$ & $10.34$ \\
$10$ & $12395949-1137230$ & $12365$ & $189.99789$ & $-11.62307$ & $298.46094$ & $51.14923$ & $0.00455$ & $0.00228$ & $1.0$ & $0$ & $0.0$ & $11.146$ & $10.906$ \\

\hline
\vspace{-2mm}
\end{tabular}
\textbf{Notes.}

a. The full catalog is available at {\tt http://gax.sjtu.edu.cn/data/Group.html}. 
\vspace{-5mm}
\label{tab_2MRS_gal}
\end{minipage}
\end{table}
\end{landscape}
}
%%%%%%%%%%%%%%%%%%%%%%%%%%%%%%%%%%%%%%%%%%%%%%%%%%%%%%%%%%%%%%%%%%%%%%%%%%%%%%

\subsubsection{The galaxy catalogs}
%%%%%HJM revised

The following items are provided for individual galaxies.

\begin{DESCRIPTION}

\item[Column (1)]
galaxy ID: unique ID of galaxies within each sample. This can 
be used to match galaxies across the galaxy and group catalogs;

\item[Column (2)]
survey ID: ID of galaxies from the original survey data release. This can 
be used to match galaxies across our catalogs and the original surveys;

\item[Column (3)]
group ID: ID of the group of which a galaxy is a member; 

\item[Column (4)]
ra (in degrees): right ascension (J2000);

\item[Column (5)]
dec (in degrees): declination (J2000);

\item[Column (6)]
$l$ (in degrees): Galactic longitude; 

\item[Column (7)]
$b$ (in degrees): Galactic latitude; 

\item[Column (8)]
$z_{\rm CMB}$: redshift in the CMB rest-frame. This is used for the group finder;

\item[Column (9)]
$z_{\rm EDD}$: redshift for nearby galaxies based on the EDD distances. Otherwise 
equals to $0$. This is only used for converting apparent magnitude to luminosity;

\item[Column (10)]
$z_{\rm comp}$: redshift completeness along the direction on the sky where 
a galaxy lies;

\item[Column (11)]
$z_{\rm src}$: a numerical value indicating the source of $z_{\rm CMB}$. 
As the sources vary for different samples, please refer to the individual catalogs 
for more detailed descriptions;

\item[Column (12)]
${\rm dist}_{\rm NN}$: angular separation to the nearest neighbor (deg) for galaxies 
that $z_{\rm src}$ is the nearest neighbor. Otherwise equals to $0$.

\item[Column (13)]
$\log (L/h^{-2}{\rm L}_\odot)$:
10-based logarithm of the luminosity in units of $h^{-2} {\rm L_\odot}$.
Luminosities are in the $K_s$-band for 2MRS and 6dFGS, in the 
$r$-band for SDSS, and in the  $R$-band for 2dFGRS. $K$- and evolutionary 
corrections to $z=0.1$ are made following \citet{lavaux11} 
(for 2MRS and 6dFGS) and \citet{poggianti97} (for SDSS and 2dFGRS).
All quantities are calculated with the assumption of the WMAP9 cosmology; 

\item[Column (14)]
$\log (M_*/h^{-2}{\rm M}_\odot)$:
10-based logarithm of the stellar mass in units of $h^{-2} {\rm M_\odot}$.
Please refer to the relevant sections for how the stellar masses are
estimated in different samples;

\item[Column (15)]
color: provided only for SDSS ($g-r$) and 2dFGRS ($b_J-R$).

\end{DESCRIPTION}

%%%%%%% SECTION 6
\section[summary]{SUMMARY}
\label{sec_summary}
%%%%HJM revised

In this paper, we have constructed group catalogs from four large redshift 
surveys in the low-$z$ universe: the 2MRS, 6dFGS, SDSS, and 2dFGRS. The groups 
are identified with a halo-based group finder that is based on the group finders 
developed in Y05, Y07 and L16 but has improved halo mass assignments
that can be applied uniformly to various observations. 
The group finder uses 
stellar mass or luminosity of central galaxies combined with the luminosity/stellar 
mass gap between the central galaxy and the $n$-th brightest/most massive satellite 
as halo mass proxies. It assigns galaxies 
into groups using halo properties, such as halo size and velocity dispersion, and
iterates with updated halo properties until the membership converges. 
We use an abundance matching technique to assign final halo masses to 
individual groups selected. For groups that are not assigned mass by abundance
matching, due to the fact that they are outside the redshift limit 
within which groups of a given mass is complete, halo masses are assigned 
based on the mean relation between halo mass and its proxy obtained from the last 
iteration of the group finder. 

We have used realistic mock galaxy samples constructed from a hydrodynamical 
simulation (EAGLE) to test the performance and to calibrate our group finder, 
and used another set of mock samples constructed from an empirical model 
of galaxy formation as an independent check. The tests showed that 
our group finder can find $\sim 95\%$ of the `true' member galaxies for about 
$95\%$ ($85\%$) of the groups for the 2MRS and 6dFGS samples
(for the SDSS and 2dFGRS samples), with better membership assignment 
for lower mass halos. The tests on mock samples also showed that 
the halo masses of individual groups estimated by the group finder are   
consistent with the true halo masses, with scatter of 
$\sim0.2\,{\rm dex}$. The scatter in the 
estimated mass - true mass relation obtained here for the SDSS sample     
is similar to Y07, but it extends uniformly to halo masses 
that are about $0.7\,{\rm dex}$ lower. 

We have constructed group catalogs by applying our group finder to the real 
redshift surveys of galaxies. From each survey, two samples of galaxies 
are constructed, one using only galaxies with spectroscopic redshifts, and 
the other using all galaxies, including  the ones with redshifts estimated 
from nearest neighbors or from photometry (photometric redshifts).
For each galaxy sample, two group catalogs are constructed, 
one using the luminosity-based halo mass proxy (Proxy-L) and the other 
using the stellar mass-based halo mass proxy (Proxy-M). 
Thus, we provide a total of $16$ group catalogs, four different sets of 
catalogs for each of the four surveys. A summary of the all the group catalogs 
and how to use them are presented in \S\ref{sec_gcatalog}.
We have also described some of the basic properties of the group catalogs, 
such as the distributions in richness, redshift,
and mass. Comparisons are made with other similar catalogs 
in the literature. 

It should be noted that the group catalogs constructed are cosmology 
dependent, and we have adopted WMAP9 cosmology in the present paper. 
This dependence comes from both the properties of dark matter halos
(halo size and velocity dispersion as functions of halo mass) 
adopted in grouping galaxies into common halos, and the halo mass function 
used in abundance matching. However, as demonstrated in Y07, 
the grouping of galaxies into groups is not sensitive to the cosmological 
model, unless the adopted model is very different from that favored
by current observations. The cosmology dependence in the halo mass 
assignments is also not a significant problem, as it is straightforward to
convert the masses to other cosmologies with abundance matching. 

The data published in this paper are available at the website:
{\tt http://gax.sjtu.edu.cn/data/Group.html}.

\section*{ACKNOWLEDGEMENTS}
%%%%HJM revised

This work is supported by the 973 Program (No. 2015CB857002) and the
national science foundation of China (grant Nos. 11233005, 11621303,
11522324, 11421303, 11503065). HJM acknowledges the support from NSF AST-1517528
and NSFC-11673015. SHL thanks his wife, Hyeji Jung, 
who took care of their kids mostly alone during writing this paper. 
We thank Hong Guo for useful discussions regarding the fiber-collision galaxies 
as well as providing the catalog of the fiber-collision galaxies in 
the SDSS DR7 that have redshift measured in the later data release,
Shiyin Shen for providing the newest LAMOST redshifts prior to their publication, 
the referee, Brent Tully, for comments that improved this paper, 
and the Virgo Consortium for making the EAGLE simulation data available. 
The EAGLE simulations were performed using the DiRAC-2 facility at Durham, 
managed by the ICC, and the PRACE facility Curie based in France at TGCC, 
CEA, Bruy\`eresle-Ch\^atel. We also acknowledge the use of the 
data products from the 2MASS, which is a joint project of the University
of Massachusetts and the Infrared Processing and Analysis
Center/California Institute of Technology, funded by the National
Aeronautics and Space Administration and the National Science
Foundation. The Wide Field Astronomy Unit at the Institute
for Astronomy, Edinburgh is acknowledged for archiving the 2MPZ catalog,
which can be accessed at http://surveys.roe.ac.uk/ssa/TWOMPZ.
Support for the development of the content for the EDD database is provided 
by the National Science Foundation under Grant No. AST09-08846.
This publication also makes use of the data products from the 2M++ catalog, 
the Final Release of the 6dFGS (available at http://www-wfau.roe.ac.uk/6dFGS/),
and the KIAS VAGC. We acknowledge the 2dFGRS team 
for the survey and for making their catalogs available, which was made 
possible through the dedicated efforts of the staff of the Anglo-Australian 
Observatory, both in creating the instrument and in supporting the survey
observations. Funding for the Sloan Digital Sky Survey IV has been provided by
the Alfred P. Sloan Foundation, the U.S. Department of Energy Office of
Science, and the Participating Institutions. SDSS-IV acknowledges
support and resources from the Center for High-Performance Computing at
the University of Utah. The SDSS web site is www.sdss.org.
SDSS-IV is managed by the Astrophysical Research Consortium for the 
Participating Institutions of the SDSS Collaboration including the 
Brazilian Participation Group, the Carnegie Institution for Science, 
Carnegie Mellon University, the Chilean Participation Group, the French 
Participation Group, Harvard-Smithsonian Center for Astrophysics, 
Instituto de Astrof\'isica de Canarias, The Johns Hopkins University, 
Kavli Institute for the Physics and Mathematics of the Universe (IPMU) / 
University of Tokyo, Lawrence Berkeley National Laboratory, 
Leibniz Institut f\"ur Astrophysik Potsdam (AIP),  
Max-Planck-Institut f\"ur Astronomie (MPIA Heidelberg), 
Max-Planck-Institut f\"ur Astrophysik (MPA Garching), 
Max-Planck-Institut f\"ur Extraterrestrische Physik (MPE), 
National Astronomical Observatories of China, New Mexico State University, 
New York University, University of Notre Dame, 
Observat\'ario Nacional / MCTI, The Ohio State University, 
Pennsylvania State University, Shanghai Astronomical Observatory, 
United Kingdom Participation Group,
Universidad Nacional Aut\'onoma de M\'exico, University of Arizona, 
University of Colorado Boulder, University of Oxford, University of Portsmouth, 
University of Utah, University of Virginia, University of Washington, 
University of Wisconsin, Vanderbilt University, and Yale University.

%\bsp

\appendix
\section{TESTING THE GROUP FINDER WITH MOCK SAMPLES CONSTRUCTED 
USING AN EMPIRICAL MODEL}
\label{sec_appendix}

%%%%%%%%%%% figure A1
\begin{figure*}
\includegraphics[width=0.75\linewidth]{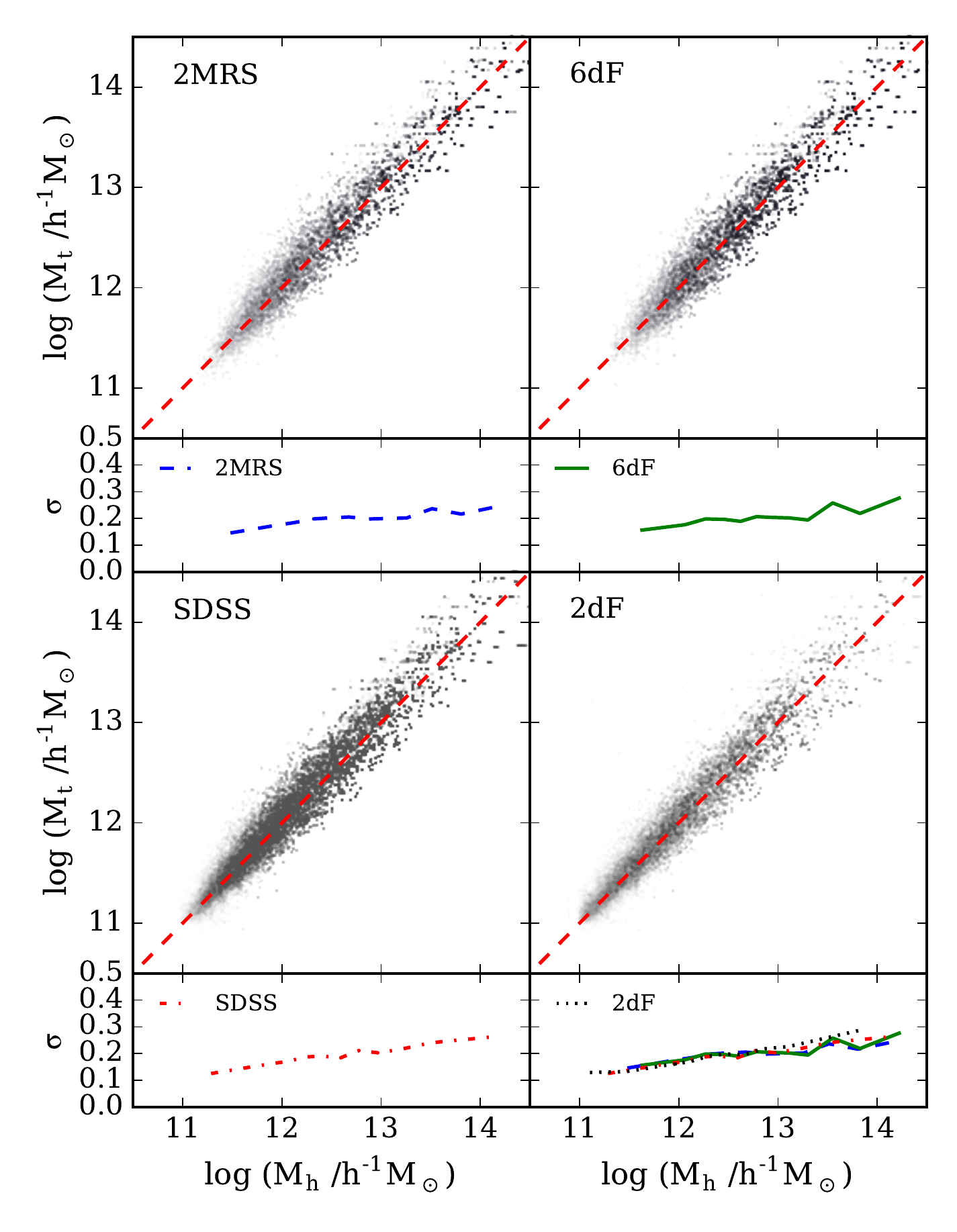}
\caption{Comparison between the true halo mass (vertical
axis) and the group mass identified by our group finder (horizontal axis) using
stellar mass as the proxy of halo mass for the mock samples of 2MRS, 6dFGS, SDSS, 
and 2dFGRS. Here mock samples are constructed by applying the 
empirical model of \citet{lu15} to the halo merger trees extracted from the 
EAGLE simulation. The small rectangular panels plot the scatter 
of the true halo masses at given group mass. }
\label{fig_mock_mass_Lu_M}
\end{figure*}

To further check the consistency of our group finder, which is calibrated 
with galaxies in the EAGLE simulation, we have applied it to 
another set of mock samples constructed using the empirical model of galaxy formation 
described in  \citet{lu15}, which is based on \citet{lu14}. The tests 
we have made are the same as those with the EAGLE mock samples presented 
in \S\ref{sec_test}, using exactly the same methods described in \S\ref{sec_gfinder}.
To do this, we first applied the empirical model to the merger trees extracted 
from the EAGLE to assign stellar masses to galaxies. As an example, 
Figure~\ref{fig_mock_mass_Lu_M} compares the true halo masses from the EAGLE
simulation with the final group masses obtained by applying our group finder 
to the mock samples thus constructed. The scatter in the halo mass comparison
is around $0.2\,{\rm dex}$, very similar to what was found in 
Figure~\ref{fig_mock_mass_M} except for the very massive end which shows  
slightly larger scatter. This demonstrates that the performance of our group finder 
is not sensitive to the details of how galaxies form in dark matter halos, 
as represented by the differences between the EAGLE and the empirical model of 
\citet{lu15}.

\label{lastpage}

\end{document}